\documentclass[
superscriptaddress,
amsmath,amssymb,
aps, 
prd,
twocolumn
]{revtex4-1}

\usepackage[dvipsnames]{xcolor}
\usepackage{graphicx}%
\usepackage{dcolumn}%
\usepackage{cases}
\usepackage{bm}%
\usepackage[normalem]{ulem}
\usepackage[colorlinks,urlcolor=Maroon,linkcolor=Maroon,anchorcolor=blue,citecolor=MidnightBlue]{hyperref} %
\allowdisplaybreaks[4]

\newcommand{\rg}{r_g}

\begin{document}

\title{\textbf{Stochastic gravitational-wave background from self-interacting superradiant clouds} 
}%

\affiliation{Department of Physics, College of Sciences, Shanghai University, 99 Shangda Road, 200444 Shanghai, China}

\author{Yin-Da Guo}
\email{yinda.guo@mail.sdu.edu.cn}
\affiliation{Key Laboratory of Particle Physics and Particle Irradiation (Ministry of Education),\\Institute of Frontier and Interdisciplinary Science, \\Shandong University, Qingdao 266237, China}
\affiliation{CENTRA, Departamento de Física, Instituto Superior Técnico – IST,\\
Universidade de Lisboa – UL, Avenida Rovisco Pais 1, 1049-001 Lisboa, Portugal}

\author{Richard~Brito}
\email{richard.brito@tecnico.ulisboa.pt}
\affiliation{CENTRA, Departamento de Física, Instituto Superior Técnico – IST,\\
Universidade de Lisboa – UL, Avenida Rovisco Pais 1, 1049-001 Lisboa, Portugal}

\author{Chen~Yuan}
\email{yuanchen@shu.edu.cn}
\affiliation{Department of Physics, College of Sciences, Shanghai University, 99 Shangda Road, 200444 Shanghai, China}
\affiliation{CENTRA, Departamento de Física, Instituto Superior Técnico – IST,\\
Universidade de Lisboa – UL, Avenida Rovisco Pais 1, 1049-001 Lisboa, Portugal}

\date{\today}%

\begin{abstract}
Gravitational-wave (GW) observations offer a powerful probe of new fundamental fields. One well-motivated source is black hole (BH)–boson cloud systems, in which an ultralight scalar field forms a cloud around a rotating BH via superradiance and emits long-lived nearly monochromatic GWs. In this work, we compute the stochastic GW background (SGWB) from such systems, extending previous work by including scalar self-interactions, and discuss its detectability with next-generation ground-based GW detectors. We find that self-interactions can suppress the SGWB and thereby relax existing LIGO-Virgo-KAGRA constraints inferred from null searches. Namely, the LIGO detectors at design sensitivity are insensitive to a SGWB produced by scalar fields with a decay constant $f_\mathrm{s} \lesssim 3\times 10^{17}$ GeV, independently of the scalar field mass. Looking ahead, we show that a moderately self-interacting cloud can still produce a detectable SGWB with next-generation detectors. Under conservative assumptions, for a decay constant $f_\mathrm{s} = 10^{17}$ GeV, the Einstein Telescope (ET) will be sensitive to scalar masses in the range $\sim[10^{-13.0},10^{-11.8}]$ eV, while Cosmic Explorer (CE) will be sensitive to scalar masses in the range $\sim[10^{-13.2},10^{-11.7}]$ eV. We also find that the minimum decay constants that still yield a detectable SGWB for ET and CE are $f_\mathrm{s}\sim 6\times10^{16}\,$GeV and $\sim 3\times10^{16}\,$GeV, respectively. 
\end{abstract}

\maketitle

\section{Introduction}
The Laser Interferometer Gravitational-Wave Observatory (LIGO) and Virgo reported the first direct detection of gravitational waves (GWs), GW150914 \cite{LIGOScientific:2016aoc}, in 2016. This observation revealed the merger of a binary black hole (BH) system, thereby inaugurating the era of GW astronomy. Since then, the growing number of observed compact-binary mergers has made GWs a powerful probe of strong-field gravity and compact objects \cite{LIGOScientific:2018mvr,LIGOScientific:2020ibl,LIGOScientific:2021usb,KAGRA:2021vkt,LIGOScientific:2025slb,LIGOScientific:2026wfs}. 

Beyond single compact binaries, the population of all undetected GW sources should also produce a stochastic GW background (SGWB), which is expected to be detected in future observing runs~\cite{LIGOScientific:2026mjf}. Binary BH systems are expected to be the main contributors to this SGWB, with a smaller contribution coming from binary neutron stars and neutron star-BH systems. However, in beyond Standard Model scenarios, additional GW sources could also contribute to the SGWB~\cite{Christensen:2018iqi}.

A prominent example is the GW emission from BH-boson cloud systems, consisting of a rotating BH surrounded by a macroscopic ``condensate'' of an ultralight bosonic field that can carry as much as $\sim10\%$ of the BH mass~\cite{East:2017ovw,Herdeiro:2017phl,Guo:2025dkx}. These systems can form because bosonic fields can extract energy and angular momentum from a spinning BH through superradiance~\cite{Zeldovich:1971ffh,Zeldovich:1972zqp,Press:1972zz}, which occurs for modes satisfying $\omega < m\,\Omega_\mathrm{H}$, where $\omega$ is the mode frequency, $m$ is the azimuthal number, and $\Omega_\mathrm{H}$ is the angular velocity of the horizon. For a massive field, the system admits quasi-bound states that are repeatedly amplified and grow exponentially~\cite{Detweiler:1980uk,Dolan:2007mj}. In the absence of interactions other than gravity, this growth continues until the BH has spun down enough that $\omega \simeq m\,\Omega_\mathrm{H}$~\cite{Brito:2014wla,East:2017ovw,Herdeiro:2017phl}. The instability is strongest when the Compton wavelength of the boson is comparable to the gravitational radius of the BH. For astrophysical BHs, this singles out ultralight bosons with masses $\lesssim 10^{-11}\,\mathrm{eV}$. For more details on BH superradiance, we refer the reader to Ref.~\cite{Brito:2015oca}.

Due to the nonaxisymmetry and time dependence of the stress-energy tensor of the cloud, the BH–cloud system emits GWs, slowly dissipating over time. Such GWs have been studied extensively in the literature, either considering
quasi-monochromatic continuous GWs emitted by single systems~\cite{Arvanitaki:2010sy,Yoshino:2013ofa,Yoshino:2014wwa,Arvanitaki:2014wva,Brito:2014wla,Arvanitaki:2016qwi,Brito:2017wnc,Brito:2017zvb,Baryakhtar:2017ngi,Isi:2018pzk,Palomba:2019vxe,Siemonsen:2019ebd,Sun:2019mqb,Brito:2020lup,Baryakhtar:2020gao,Zhu:2020tht,LIGOScientific:2021rnv,KAGRA:2022osp,Collaviti:2024mvh,Jones:2024fpg,Guo:2024dqd,Guo:2025dkx,Mirasola:2025car,Ellis:2026gwt}, considering the SGWB emitted by a large population of BH-cloud systems \cite{Brito:2017wnc,Brito:2017zvb,Tsukada:2018mbp,Tsukada:2020lgt,Zhu:2020tht,Yuan:2021ebu,Yuan:2022bem,Guo:2023gfc,Ellis:2026gwt}, or considering level-transition emission \cite{Arvanitaki:2010sy,Arvanitaki:2014wva,Baryakhtar:2020gao,Omiya:2024xlz,Lyu:2026mbv} and interference-induced GW ``beats'' \cite{Siemonsen:2019ebd,Guo:2022mpr,Guo:2024dqd,Guo:2025dkx} when more than one mode is present in the cloud. %
Blind searches as well as targeted searches have been conducted for both continuous GWs and the SGWB, considering not only ultralight scalar fields but also ultralight vector fields, and, so far, only null results have been reported \cite{Palomba:2019vxe,LIGOScientific:2021rnv,KAGRA:2022osp,Tsukada:2018mbp,Tsukada:2020lgt,Sun:2019mqb,Yuan:2022bem,Guo:2023gfc,LIGOScientific:2026pwx}. Therefore, these searches already place constraints on ultralight boson masses in a range roughly given by $\sim[10^{-13},10^{-12}]\mathrm{eV}$.

Complementary constraints arise indirectly from BH spin measurements. The formation of the BH–cloud system is typically accompanied by the spin-down of the BH, leading to a gap in the Regge plane, i.e., the plane spanned by BH mass and spin~\cite{Arvanitaki:2010sy}. Consequently, measurements of stellar mass BH populations can constrain ultralight bosons in a similar mass band \cite{Arvanitaki:2010sy,Cardoso:2018tly,Fernandez:2019qbj,Ng:2019jsx,Ng:2020ruv,Cheng:2022jsw,Guo:2024dqd,Hoof:2024quk,Aswathi:2025nxa,Ning:2026ebu}. There have been claims that a scalar boson with mass $\sim 10^{-12}$ eV explains the spin distribution of the binary BHs observed with LIGO-Virgo-KAGRA (LVK)~\cite{Kou:2026naz}, although this claim seems inconsistent with constraints already coming from BH spin measurements in X-ray binaries~\cite{Arvanitaki:2014wva,Witte:2024drg}.  

However, an important caveat in both direct and indirect constraints is that they can be relaxed in parts of parameter space when self-interactions are non-negligible~\cite{Baryakhtar:2020gao,Witte:2024drg}. Except for few exceptions~\cite{Collaviti:2024mvh}, most studies related to the detectability of the GW emission from boson clouds neglect self-interactions, which is justified when the relevant decay constant is close to the Planck scale, so that the self-couplings are sufficiently weak. If only gravitational interactions are included, the cloud evolution can be divided into two stages: one dominated by superradiance and the other dominated by GW emission. When self-interactions become appreciable, an additional evolutionary stage appears that is dominated by self-interaction processes \cite{Baryakhtar:2020gao,Omiya:2022mwv,Omiya:2022gwu,Collaviti:2024mvh,Witte:2024drg}. This extra stage can suppress the occupation numbers of bosons in the cloud and shorten the GW emission stage, potentially weakening continuous GW signals and the SGWB in parts of parameter space. Moreover, because the occupation numbers are suppressed, the rate at which an unstable BH spins down decreases when self-interactions are sufficiently strong. Therefore, the existence of rapidly spinning BHs need not be in tension with ultralight bosons, and the corresponding indirect constraints may also be relaxed~\cite{Arvanitaki:2014wva,Hoof:2024quk,Witte:2024drg} for large enough self-couplings. These effects motivate a dedicated, quantitative reassessment of SGWB predictions once self-interactions are included. In this work, we close this gap by studying how self-interactions affect the SGWB from BH-cloud systems, focusing on ultralight scalar fields.

This paper is organized as follows. In Sec.~\ref{sec:superradiance}, we briefly review quasi-bound states in BH spacetimes and describe how we incorporate self-interactions. In Sec.~\ref{sec:evo}, we study the evolution of the BH-cloud system in the presence of self-interactions and outline how we compute the total energy emitted in GWs by a single system. In Sec.~\ref{sec:SGWB}, we present the resulting SGWB energy density spectrum and its signal-to-noise ratio (SNR) for different detectors, and discuss its detectability and the prospects for constraints with next-generation detectors. We summarize our results and conclude in Sec.~\ref{sec:Summary}. Throughout the paper, we adopt the natural unit system $\hbar=c=1$.

\section{Scalar field in a Kerr black hole background}
\label{sec:superradiance}

The Kerr metric describes a rotating BH characterized by its mass $M$ and angular momentum $J$. In Boyer-Lindquist coordinates, it can be written as \cite{Boyer:1966qh}:
\begin{align}
    \begin{split}
        ds^2=&-\left( 1-\frac{2 \rg r}{\Sigma} \right) dt^2-\frac{4a \rg r}{\Sigma}\sin ^2\theta dtd\varphi +\frac{\Sigma}{\Delta}dr^2
        \\
        &+\Sigma d\theta ^2+\left[ \left( r^2+a^2 \right) \sin ^2\theta +2\frac{\rg r}{\Sigma}a^2\sin ^4\theta \right] d\varphi ^2,
    \label{eq:KerrMetric}
    \end{split}
\end{align}
where
\begin{equation}
    \Delta  \equiv r^2-2 \rg r+a^2,\quad
    \Sigma  \equiv r^2+a^2 \cos ^2 \theta\,,
\end{equation}
and $a \equiv J/M$ is the BH's angular momentum per unit mass, while $\rg \equiv GM$.
Here, $G$ is the gravitational constant, and the Planck mass is defined as $M_\mathrm{pl}\equiv1/\sqrt{G}$. For later use it is also useful to define the dimensionless spin parameter $a_* \equiv a/\rg$. The BH possesses an inner horizon $r_-$ and an outer horizon $r_+$, located at
\begin{align}
    r_{ \pm}=\rg \pm \sqrt{\rg^2-a^2}\,,
\end{align}
with the requirement that $0\leq a_*\leq 1$ for the existence of an event horizon.

We now consider a real scalar field $\Phi$ propagating in a Kerr BH background. The corresponding Lagrangian density is
\begin{align}\label{eq:Lagrangian}
  \mathcal{L} = - \frac{1}{2}\nabla^{\mu}\Phi \nabla_{\mu}\Phi - V (\Phi)\,.
\end{align}
Including self-interactions and neglecting interactions with other fields, we consider a potential term given by
\begin{align}
  V (\Phi) = \frac{1}{2} \mu_\mathrm{s}^2 \Phi^2 - \frac{\lambda}{4!}\Phi^4+\mathcal{O}(\Phi^6).
\end{align}
where $\mu_\mathrm{s}$ denotes the mass of the scalar field and $\lambda$ is the dimensionless self-interaction coupling parameter. For axion-like particles this coupling parameter can be related to a dimensionful decay constant $f_\mathrm{s}$ as \cite{Baryakhtar:2020gao}
\begin{align}
  \lambda \equiv \frac{\mu_\mathrm{s}^2}{f_\mathrm{s}^2} \simeq 10^{-74}\left(\frac{\mu_\mathrm{s}}{10^{-12}\,\mathrm{eV}}\right)^2\left(\frac{10^{16}\,\mathrm{GeV}}{f_\mathrm{s}}\right)^2.
\end{align}

This Lagrangian density leads to the equation of motion
\begin{align}\label{eq:EoM}
  (\nabla^{\mu} \nabla_{\mu} - \mu_\mathrm{s}^2) \Phi = -\frac{\lambda}{3!}\Phi^3.
\end{align}
In the limit $|\Phi/f_\mathrm{s}|\ll1$, the cubic term can be neglected and the equation of motion reduces to the Klein-Gordon equation,
\begin{align}\label{eq:EoM_0th}
  (\nabla^{\mu} \nabla_{\mu} - \mu_\mathrm{s}^2) \Phi = 0.
\end{align}
Imposing quasi-bound state boundary conditions for the Klein-Gordon equation, i.e. ingoing waves at the BH's outer horizon and an exponentially decaying field at spatial infinity, the system admits solutions corresponding to quasi-bound states with complex eigenfrequencies~\cite{Detweiler:1980uk,Dolan:2007mj}, which we write as $\omega_{nlm}+i \Gamma_{nlm}$ for $\omega_{nlm}$ and $\Gamma_{nlm}$ the real and imaginary parts of the eigenfrequencies, respectively. The eigenfrequencies can be labelled by three quantum numbers, $n$, $l$ and $m$, which denote the overtone, angular, and azimuthal numbers, respectively. In the non-relativistic regime, $\alpha \equiv \rg\mu_\mathrm{s} \ll 1$, the real part $\omega_{nlm}$ admits a power-series expansion in $\alpha$~\cite{Baumann:2019eav}
\begin{align}\label{eq:omega}
	\omega_{nlm} &\approx \mu_\mathrm{s} \Big(1-\frac{\alpha^2}{2\bar{n}^2}-\frac{\alpha^4}{8\bar{n}^4}+\frac{f_{\bar{n}l}}{\bar{n}^3}\alpha^4+\frac{h_l a_* m}{\bar{n}^3}\alpha^5\Big),
\end{align}
where we introduced the principal number $\bar{n} = n + l + 1$, and defined
\begin{align}
	f_{\bar{n}l} & \equiv -\frac{6}{2l+1}+\frac{2}{\bar{n}},\\
	h_{l} & \equiv \frac{16}{2l(2l+1)(2l+2)}.
\end{align}
In the same limit, the imaginary part $\Gamma_{nlm}$ can be written as \cite{Bao:2022hew,Bao:2023xna,Guo:2025dkx}:
\begin{align}\label{eq:Gamma}
	\begin{split}
		& \Gamma_{nlm} \approx
    \\
		& \hspace{0.3cm} -\omega_1 \left(4\kappa\sqrt{\rg^2-a^2}\right)^{2l^\prime+1}\frac{\Gamma (n + 2 l^\prime + 2)}{n!}\frac{\sinh (2 \pi  p)}{2 \pi }
		\\
		& \hspace{-0.2cm} \times\frac{\left| \Gamma \left(l^\prime+1-i p+\sqrt{q-p^2}\right) \Gamma \left(l^\prime+1+i p+\sqrt{q-p^2}\right) \right| ^2}{\left[\Gamma ( 2 l^\prime + 1)\Gamma ( 2 l^\prime + 2)\right]^2},
	\end{split}
\end{align}
where $l^\prime \equiv l+\epsilon$, $p \equiv \rg r_+ (\omega_{nlm} - m \Omega_\text{H}) / \sqrt{\rg^2-a^2}$, $\kappa \equiv \sqrt{\mu_\mathrm{s}^2-\omega_{0}^2}$, $ \Omega_\mathrm{H} \equiv a/(2\rg r_+)$, and
\begin{subequations}
\begin{align}
	\epsilon & \equiv -\frac{8\alpha^2}{2l+1},
	\\
  \begin{split}
    q & \equiv \frac{8 \rg r_+ \omega_{nlm} (r_+\omega_{nlm} - m \rg \Omega_\text{H})}{r_+ - r_-} 
    \\
    & \hspace{1cm} - \mu_\mathrm{s}^2 (r_+^2 + a^2) +4 \rg^2 (\mu_\mathrm{s}^2 - 3 \omega^2_{nlm}),
  \end{split}
	\\
	\omega_0 & \equiv \mu_\mathrm{s} \sqrt{1 - \frac{2 \alpha^2}{\bar{n}^2 +4\alpha^2 + \bar{n} \sqrt{\bar{n}^2 + 8 \alpha^2}}},
	\\
	\omega_1 & \equiv \frac{\mu_\mathrm{s}^2 - \omega_{0}^2}{\bar{n} \omega_{0} (1 + 4 \rg^2 (2\omega_{0}^2 - \mu_\mathrm{s}^2) / \bar{n}^2)}.
\end{align}
\end{subequations}
From this expression, one can see that there is a threshold frequency at which $\Gamma_{nlm}=0$, namely $\omega_{nlm}=m\Omega_\mathrm{H}$. For $\omega_{nlm}<m\Omega_\mathrm{H}$ one has $\Gamma_{nlm}>0$ and the modes are unstable, with modes growing exponentially on an e-folding timescale $1/\Gamma_{nlm}$, while for $\omega_{nlm}>m\Omega_\mathrm{H}$, $\Gamma_{nlm}<0$ and the modes decay exponentially on an e-folding timescale $1/|\Gamma_{nlm}|$. The corresponding critical BH spin where $\omega_{nlm}=m\Omega_\mathrm{H}$ is given by
\begin{align}\label{eq:asc}
	a_{\mathrm{*c},nlm} = \frac{4m \rg \omega_{nlm}}{m^2+(2\rg\omega_{nlm})^2}.
\end{align}

In what follows we use the approach of Ref.~\cite{Baryakhtar:2020gao}, where one solves the equation of motion~\eqref{eq:EoM} perturbatively taking $|\Phi/f_\mathrm{s}|\ll1$, and also assuming a small $\alpha\ll 1$ approximation. At leading order the solution is then given by quasi-bound state solutions of the Klein-Gordon equation in Kerr, while higher-order corrections lead to the emission of scalar radiation due to the cubic term in Eq.~\eqref{eq:EoM}. The resulting scalar fluxes from this calculation are listed in the next section. For details of the perturbative solution of Eq.~\eqref{eq:EoM}, we refer the reader to App.~B of Ref.~\cite{Baryakhtar:2020gao}.

\section{Evolution of a scalar cloud with self-interactions}
\label{sec:evo}
In this section, we first derive the evolution equations of the BH-cloud system in Sec.~\ref{sec:evo_eqs}, following Ref.~\cite{Baryakhtar:2020gao}. We then present the numerical results and review the analytical formulas describing the evolution in Sec.~\ref{sec:Numerical_results}. Finally, in Sec.~\ref{sec:Total_GW}, we outline the procedure used to compute the total GW emission energy, which we will use in the next section to estimate the SGWB from a population of BH-cloud systems.

\subsection{Evolution equations}
\label{sec:evo_eqs}

Based on a small $\alpha$ approximation, Ref.~\cite{Baryakhtar:2020gao} argued that for $\alpha \lesssim 0.22$ and with the $\{0,1,1\}$ mode initially growing due to the superradiant instability, the scalar cloud evolution can be approximately described as a closed two-mode system consisting of the $\{0,1,1\}$ and $\{0,2,2\}$ modes. For the sake of keeping the analysis simple, here we will use the two-mode description, and check a posteriori that the approximation should not strongly affect our main conclusions.

To describe the evolution of the BH-cloud system in the presence of self-interactions, several physical processes need to be taken into account. First, the scalar cloud can extract energy and angular momentum from the BH through the superradiant mechanism, or transfer energy and angular momentum back to the BH when the superradiant condition is not satisfied. This process can be schematically represented as
\begin{align}
	\mathrm{BH} \longleftrightarrow nlm.
\end{align}
The corresponding energy and angular momentum fluxes through the BH horizon that describe this process are given by 
\begin{align}
	\dot{E}_{\mathrm{BH}} &= - \sum_{nlm} 2 \omega_{nlm} N_{nlm}\Gamma_{nlm},
    \\
    \dot{J}_{\mathrm{BH}} &= - \sum_{nlm} 2 m N_{nlm}\Gamma_{nlm},
\end{align}
where $N_{nlm}$ denotes the occupation number of the scalar cloud in a single $\{nlm\}$ mode, and we used the fact that each scalar quantum carries energy $\omega_{nlm}$ and angular momentum $m$.

The other gravitational processes involve GW emission, namely due to scalar annihilation or transition between scalar energy states~\cite{Arvanitaki:2014wva}. 
Given the two-mode approximation, for the annihilation process, there are three possible channels:
\begin{align}
	011\,\times\,011 &\rightarrow \mathrm{GW},\label{GW_anni_011}
    \\
    011\,\times\,022 &\rightarrow \mathrm{GW},\label{GW_anni_011022}
    \\
    022\,\times\,022 &\rightarrow \mathrm{GW}.
\end{align}
To leading order in $\alpha$, the GW luminosities corresponding to each process are given by
\begin{align}
	\dot{E}_{011\times011}^{\mathrm{GW}} &\approx \frac{484+9\pi^2}{23040} \frac{1}{G}\varepsilon_{011}^2\alpha^{16},
    \\
    \dot E_{011\times022}^{\rm GW} &\approx \frac{256+9\pi^2}{8817984} \frac{1}{G}\varepsilon_{011}\varepsilon_{022}\alpha^{18},
    \\
    \dot{E}_{022\times022}^{\mathrm{GW}} &\approx \frac{1024+49\pi^2}{5423886846} \frac{1}{G}\varepsilon_{022}^2\alpha^{20},
\end{align}
where $\varepsilon_{nlm}=N_{nlm}/(\rg M)$ is the normalized occupation number. Here and below, the subscripts in $\dot{E}$ and $\dot{J}$, as well as the coefficient $\gamma$ used later, denote the initial state of the corresponding process, while superscripts denote the final state. The self-annihilation luminosities are taken from Refs.~\cite{Brito:2014wla,Guo:2022mpr}, whereas the mixed-channel luminosity is derived in this work using the same framework. The GW frequency for any of these annihilation processes can be approximated as $\widetilde{\omega}_\mathrm{ann}\approx 2\mu_\mathrm{s}$.

On the other hand, for the transition process, there is a single allowed channel:
\begin{align}
	022 \rightarrow 011\,\times\,\mathrm{GW}.
\end{align}
The corresponding GW emission flux is given by \cite{Arvanitaki:2010sy,Baryakhtar:2020gao}
\begin{align}
	\dot{E}_{022}^{011\times\mathrm{GW}} \approx \frac{2^8\times5717}{3^5 5^{11} 7^3} \frac{1}{G} \varepsilon_{011} \varepsilon_{022} \alpha^{14}\,.
\end{align}
The GW frequency of this transition process can be approximated as $\widetilde{\omega}_\mathrm{trans}\approx \omega_{022}-\omega_{011}\approx \frac{5}{72} \alpha^2 \mu_\mathrm{s}$.

Besides purely gravitational processes, self-interactions induce additional energy and angular momentum losses, which can be classified into three types. The first type consists of relativistic scalar emission towards infinity due to the ``annihilation'' of three scalar quanta in the two different states, producing relativistic scalars with energy $\omega\approx3\mu_\mathrm{s}$. The possible interactions in this case are:
\begin{align}
    011\,\times\,011\,\times\,011\, &\rightarrow\, \infty,
    \\
    011\,\times\,011\,\times\,022\, &\rightarrow\, \infty,
    \\
    011\,\times\,022\,\times\,022\, &\rightarrow\, \infty,
    \\
    022\,\times\,022\,\times\,022\, &\rightarrow\, \infty.
\end{align}
The scalar emission flux for these processes is given by \cite{Baryakhtar:2020gao}
\begin{align}
    \begin{split}
        \dot{E}_{nlm\times n'l'm'\times n''l''m''}^{\infty} &\propto 
        \\
        & \hspace{-1cm} \alpha^{11+2(l+l'+l'')}\mu^2_\mathrm{s}N_{nlm}N_{n'l'm'}N_{n''l''m''}.
    \end{split}
\end{align}

A second type of interactions leads instead to non-relativistic scalar emission, namely through the process 
\begin{align}
    022\,\times\,022\, \rightarrow\, 011\,\times\, \infty\,,
\end{align}
which can be interpreted as the ``annihilation'' of two $\{0,2,2\}$-mode scalars producing a $\{0,1,1\}$-mode scalar and an emitted non-relativistic scalar with energy $\omega\approx\mu_\mathrm{s}(1+\alpha^2/72)$. The corresponding scalar emission flux is \cite{Baryakhtar:2020gao}
\begin{align}
    \dot{E}_{022\times 022}^{011\times\infty} = 10^{-8} \alpha^4 \lambda^2 \mu_\mathrm{s}^2 N_{022}^2N_{011}.
\end{align}

Finally, the third type of possible relevant mechanism is a process in which the ``annihilation'' of two $\{0,1,1\}$-mode scalars produces a $\{0,2,2\}$-mode scalar as well as a bound forced oscillation with frequency $\omega\approx\mu_\mathrm{s}(1-7\alpha^2/36)$, which is absorbed by the BH:
\begin{align}
    011\,\times\,011\, &\rightarrow\, 022\,\times\, \mathrm{BH}.
\end{align}
The energy flux carried through the BH horizon in this process is \cite{Baryakhtar:2020gao}
\begin{align}
    \dot{E}_{011\times 011}^{022\times\mathrm{BH}} = 4\times10^{-7} \alpha^7 \lambda^2 (1+\sqrt{1-a_*^2})\mu_\mathrm{s}^2 N_{011}^2N_{022}.
\end{align}

In general, since the relativistic scalar emission scales with a higher power in $\alpha$ than the other two types of processes, this emission channel can be neglected for the purposes of understanding the overall evolution of the cloud~\cite{Baryakhtar:2020gao}. 

Using conservation of energy and angular momentum, we can then obtain the equations describing the evolution of the two-mode model~\cite{Baryakhtar:2020gao,Collaviti:2024mvh}:
\begin{subequations}\label{eq:evo_eqs}
\begin{align}
    \begin{split}
        \dot{\varepsilon}_{011} +\frac{2\dot{M}}{M}\varepsilon_{011} 
        &= \gamma_{\mathrm{BH}}^{011}\varepsilon_{011} - 2\gamma_{011\times011}^{\mathrm{GW}}\varepsilon_{011}^2 
        \\
        &-\gamma_{011\times022}^{\rm GW} \varepsilon_{011}\varepsilon_{022} + \gamma_{022}^{011\times\mathrm{GW}}\varepsilon_{011} \varepsilon_{022}
        \\
        &-2\gamma^{022\times\mathrm{BH}}_{011\times011}\varepsilon_{011}^2\varepsilon_{022} + \gamma^{011 \times \infty}_{022 \times 022} \varepsilon_{011} \varepsilon_{022}^{2},
    \end{split}
    \\
    \begin{split}
        \dot{\varepsilon}_{022} +\frac{2\dot{M}}{M}\varepsilon_{022} 
        &= \gamma_{\mathrm{BH}}^{022}\varepsilon_{022} - 2\gamma_{022\times022}^{\mathrm{GW}}\varepsilon_{022}^2 \\
        &-\gamma_{011\times022}^{\rm GW} \varepsilon_{011}\varepsilon_{022} - \gamma_{022}^{011\times\mathrm{GW}}\varepsilon_{011} \varepsilon_{022}
        \\
        &+ \gamma^{022 \times \mathrm{BH}}_{011 \times 011} \varepsilon_{011}^{2} \varepsilon_{022} - 2 \gamma^{011 \times \infty}_{022 \times 022} \varepsilon_{011} \varepsilon_{022}^{2}
        ,
    \end{split}
    \\
    \begin{split}
        \dot{a}_*+\frac{2\dot{M}}{M}a_*
        &= - \gamma^{011}_{\mathrm{BH}} \varepsilon_{011} - 2 \gamma^{022}_{\mathrm{BH}} \varepsilon_{022},
    \end{split}
    \\
    \begin{split}
        \frac{\dot{M}}{G M^2} &= - \omega_{011} \gamma^{011}_{\mathrm{BH}} \varepsilon_{011} - \omega_{022} \gamma^{022}_{\mathrm{BH}} \varepsilon_{022} 
        \\
        & \hspace{0.4cm} + (2\omega_{011}-\omega_{022}) \gamma^{022 \times \mathrm{BH}}_{011 \times 011} \varepsilon_{011}^{2} \varepsilon_{022},
    \end{split}
\end{align}
\end{subequations}
where we defined
\begin{align}
    \gamma_{\mathrm{BH}}^{nlm} &= 2\Gamma_{nlm},
    \\
    \gamma_{011\times011}^{\mathrm{GW}} &=\frac{1}{2}\frac{484+9\pi^2}{23040} \alpha^{14}\mu_\mathrm{s},
    \\
    \gamma_{011\times022}^{\rm GW} &= \frac{256+9\pi^2}{17635968}\alpha^{16}\mu_\mathrm{s},
    \\
    \gamma_{022\times022}^{\mathrm{GW}} &= \frac{1}{2}\frac{1024+49\pi^2}{5423886846} \alpha^{18} \mu_\mathrm{s}, 
    \\
    \gamma_{022}^{011\times\mathrm{GW}} &= \frac{72}{5}\frac{2^8\times5717}{3^5 5^{11} 7^3} \alpha^{10} \mu_\mathrm{s},
    \\
    \gamma^{022 \times \mathrm{BH}}_{011 \times 011} &= 4\times10^{-7} \alpha^{11} \left(\frac{M_\mathrm{pl}}{f_\mathrm{s}}\right)^4(1+\sqrt{1-a_*^2})\mu_\mathrm{s},
    \\
    \gamma^{011 \times \infty}_{022 \times 022} &= 10^{-8} \alpha^8 \left(\frac{M_\mathrm{pl}}{f_\mathrm{s}}\right)^4 \mu_\mathrm{s}.
\end{align}

\subsection{Numerical results and analytical approximations}
\label{sec:Numerical_results}

\begin{figure*}
    \centering
    \includegraphics[width=0.99\linewidth]{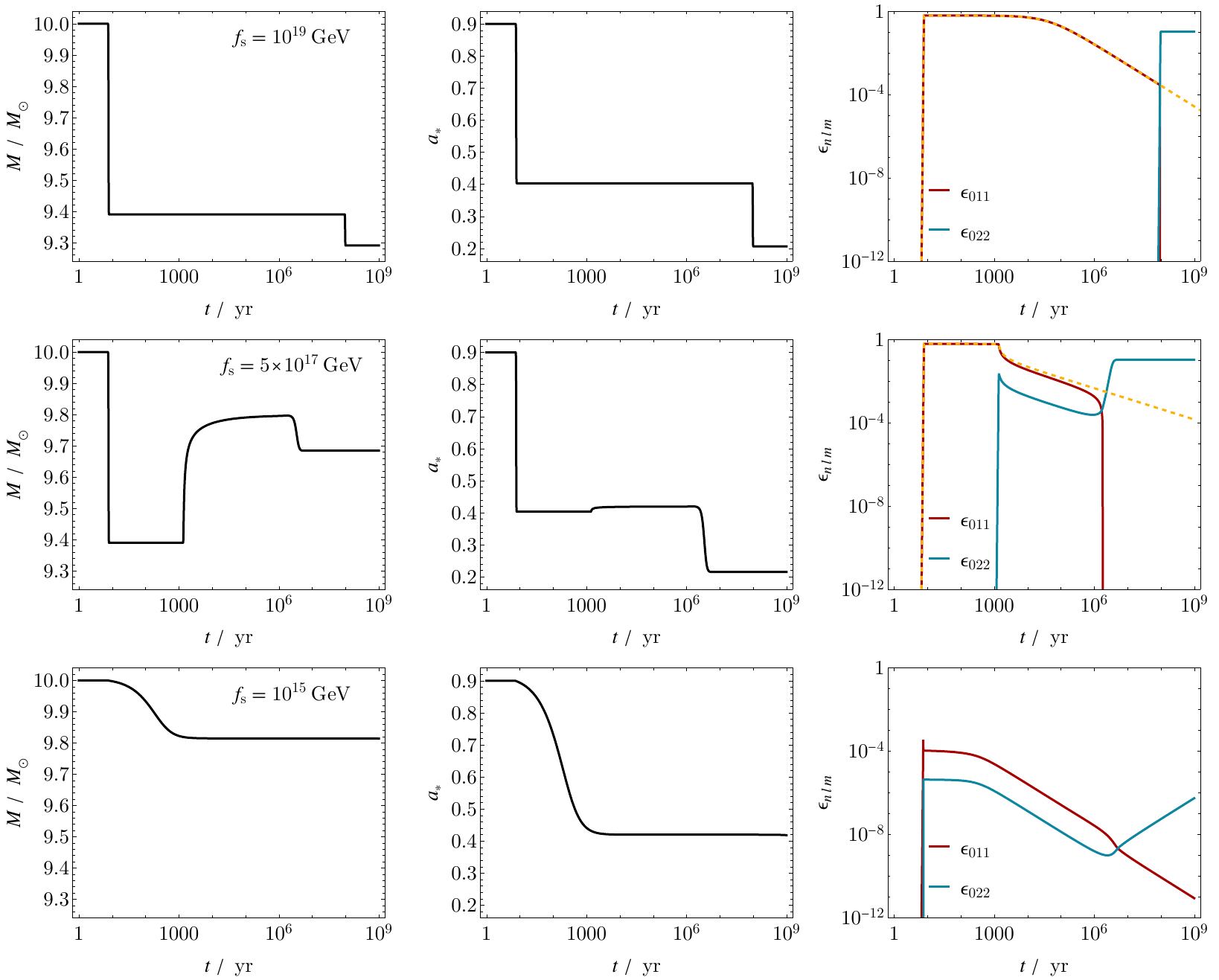}
    \caption{The evolution of BH mass, BH spin, and the scalar normalized occupation numbers as a function of time (measured in years). The solid curves denote the numerical results of Eqs.~\eqref{eq:evo_eqs}. The orange dashed curves represent the normalized occupation number of the $\{0,1,1\}$ mode computed from the analytical formulas presented in the text. The initial BH mass, initial BH spin, and scalar mass are set to $M_0 = 10 M_\odot$, $a_{*0} = 0.9$, and $\mu_\mathrm{s} = 1.5\times10^{-12}$~eV, respectively. The decay constant is taken to be $f_\mathrm{s} = 10^{19}$~GeV for the first row, $f_\mathrm{s} = 5\times10^{17}$~GeV for the second row, and $f_\mathrm{s} = 10^{15}$~GeV for the third row.}
    \label{fig:evo}
\end{figure*}

The set of Eqs.~\eqref{eq:evo_eqs} can be solved numerically. We show examples of such evolutions in Fig.~\ref{fig:evo}, where the results of the numerical solutions are shown with solid lines. For this set of simulations, we set the initial BH mass, initial BH spin, and scalar mass to $M_0 = 10 M_\odot$, $a_{*0} = 0.9$, and $\mu_\mathrm{s} = 1.5\times10^{-12}$~eV, respectively. The corresponding dimensionless initial mass coupling is $\alpha_0 \approx 0.112$. To model initial seeding by quantum fluctuations, we initialize both modes with $N_{011}(0)=N_{022}(0)=1$, corresponding to $\varepsilon_{011,0}=\varepsilon_{022,0}=1/(GM_0^2)$. We compare results obtained with three different decay constants: $f_\mathrm{s} = 10^{19}$~GeV (first row), $f_\mathrm{s} = 5\times10^{17}$~GeV (second row), and $f_\mathrm{s} = 10^{15}$~GeV (third row). To study the GW energy emitted by the $\{0,1,1\}$ mode, we also plot the normalized occupation number of this mode as orange dashed curves, computed using the analytical formulas introduced below.

Depending on the decay constant $f_\mathrm{s}$, the evolution of the BH-cloud system can be classified into three cases: {\it small self-coupling regime}, {\it moderate self-coupling regime}, and {\it large self-coupling regime} \cite{Baryakhtar:2020gao}. The small self-coupling regime occurs when \cite{Baryakhtar:2020gao}
\begin{align}
    \begin{split}
        &f_\mathrm{s} > f_{\mathrm{s},1} \approx \\
        &\min\Bigg[3\times10^{16}\,\mathrm{GeV} \left(\frac{T_\mathrm{BH}}{10^{10}\,\mathrm{yr}}\right)^{\frac14}\left(\frac{\mu_\mathrm{s}}{10^{-13}\,\mathrm{eV}}\right)^{\frac14}\left(\frac{\alpha}{0.01}\right)^{\frac{11}{4}},
        \\
        &\hspace{0.9cm}8\times10^{18}\,\mathrm{GeV} \left(\frac{0.01}{\alpha}\right)^{\frac{3}{4}}\left(\frac{a_*}{0.9}\right)^{\frac14} \Bigg],
    \end{split}
\end{align}
where $T_\mathrm{BH}$ is the lifetime of the BH. Substituting the parameters used in Fig.~\ref{fig:evo}, we obtain $f_{\mathrm{s},1} \approx 1.30\times10^{18}$ GeV. Therefore, the first row of Fig.~\ref{fig:evo} belongs to the small self-coupling regime. 

In the small self-coupling regime, the evolution of the system can be regarded as purely gravitational, with self-interactions being negligible. Initially, the $\{0,1,1\}$-mode scalar is seeded by quantum fluctuations and subsequently grows exponentially due to superradiance. Meanwhile, energy and angular momentum are extracted from the BH and transferred to the scalar cloud until the BH spins down to the critical value $a_{*\mathrm{c}}$, with the corresponding time denoted by $t_1$. At the same time, the BH mass decreases to \cite{Tsukada:2018mbp}
\begin{align}
    M(t_1)=\frac{1-\sqrt{1-16\omega_{011}^2 r_{g,0}^2(1-\omega_{011} r_{g,0} a_{*0})^2}}{8\omega_{011}^2 r_{g,0}^2(1-\omega_{011} r_{g,0} a_{*0})} M_0,
\end{align}
where $r_{g,0}\equiv GM_0$. For Fig.~\ref{fig:evo}, the estimated value is 9.38962 $M_\odot$, to be compared with the numerical value of 9.38963 $M_\odot$. The mass of the $\{0,1,1\}$ mode reaches its maximum at $t_1$ when $M_\mathrm{s,011}(t_1) = M_0 - M(t_1)$, and the corresponding normalized occupation number can be estimated as 
\begin{align}
    \varepsilon_{011}(t_1) = \frac{N_{011}(t_1)}{GM(t_1)^2} = \frac{M_\mathrm{s,011}(t_1)/M(t_1)}{GM(t_1)\omega_{011}}.
\end{align}
For Fig.~\ref{fig:evo}, the estimated value is 0.61763, to be compared with the numerical value of 0.61766. We can then estimate the timescale $t_1$ using
\begin{align}
    t_1 = \frac{1}{\gamma^{011}_\mathrm{BH}}\log\frac{\varepsilon_{011}(t_1)}{\varepsilon_{011,0}}.
\end{align}

After $t_1$, the superradiance process of the $\{0,1,1\}$ mode shuts off and GW emission due to the $011\times011\rightarrow\mathrm{GW}$ annihilation process dominates the evolution of the system. During this phase, both the BH mass and spin remain approximately constant up to the point where the superradiant instability of the $m=2$ mode starts becoming important. On the other hand, the normalized occupation number of the $\{0,1,1\}$ mode decreases due to GW emission, decaying approximately as \cite{Brito:2017zvb}
\begin{align}\label{eq:eps_after_t1}
    \varepsilon_{011}(t>t_1) = \frac{\varepsilon_{011}(t_1)}{1+(t-t_1)/\tau_\mathrm{GW,011}},
\end{align}
with 
\begin{align}
   \tau_\mathrm{GW,011} \equiv \frac{1}{2\varepsilon_{011}(t_1)\gamma^{\mathrm{GW}}_{011\times011}}.
\end{align}
Accordingly, the total GW emission energy over the interval $[t_1,t]$ is given by
\begin{align}\label{eq:EGW_total_1}
   E_\mathrm{GW}(t) \approx \frac{M_\mathrm{s,011}(t_1) (t-t_1)}{t-t_1+\tau_\mathrm{GW,011}}\,.
\end{align}
Given that GW emission before $t=t_1$ is extremely suppressed, this can also be considered to be the total energy emitted since $t\sim 0$.

Finally, the $m>1$ modes undergo an analogous sequence of evolution. In this work, we focus on the GWs generated by the $\{0,1,1\}$ mode. For the SGWB, this provides a good approximation for $\mu_\mathrm{s}\lesssim 10^{-12}$~eV \cite{Yuan:2021ebu}.

The moderate self-coupling regime corresponds to \cite{Baryakhtar:2020gao}
\begin{align}\label{eq:f2}
    \begin{split}
        &f_\mathrm{s,1} > f_\mathrm{s} > f_\mathrm{s,2} \approx
        \\
        &2\times10^{16}\,\mathrm{GeV} \left(\frac{a_{*0}}{0.9}\right)^{\frac{1}{4}}\min\left[\left(\frac{\alpha}{0.04}\right)^{\frac{3}{4}},\left(\frac{\alpha}{0.04}\right)^{\frac{3}{2}}\right]
    \end{split}
\end{align}
Substituting the parameters adopted in Fig.~\ref{fig:evo}, we obtain $f_{\mathrm{s},2} \approx 4.34\times10^{16}$~GeV. Accordingly, the second row of Fig.~\ref{fig:evo} belongs to the moderate self-coupling regime. In this regime, the early growth of the $\{0,2,2\}$ mode occurs, driven by the $011\times011\rightarrow022\times\mathrm{BH}$ process. The corresponding time is denoted by $t_2$, which can be estimated as \cite{Baryakhtar:2020gao}
\begin{align}\label{eq:t_2}
    t_2 \approx t_1 + \frac{\log[GM(t_1)^2]}{\gamma^{022\times\mathrm{BH}}_{011\times011}\varepsilon_{011}(t_1)^2}.
\end{align}
For $t<t_2$, the evolution of the system is similar to that in the small self-coupling regime, where gravitational processes dominate. Note that the estimate of $t_2$ obtained from Eq.~\eqref{eq:t_2} becomes inaccurate when $\varepsilon_{011}(t_2) \ll \varepsilon_{011}(t_1)$ due to GW emission. In practice, we initialize $t_2$ using Eq.~\eqref{eq:t_2}, update $\varepsilon_{011}$ at the estimated time using Eq.~\eqref{eq:eps_after_t1}, and insert this value back into Eq.~\eqref{eq:t_2}. We perform three such updates, holding the BH parameters and rate coefficients fixed at their values at $t_1$.

After $t_2$, the self-interaction processes $011\times011\rightarrow022\times\mathrm{BH}$ and $022\times022\rightarrow011\times\infty$ dominate the evolution and drive the system toward an equilibrium. Since the energy is transferred to the BH and to infinity, the occupation numbers of both the $\{0,1,1\}$ and $\{0,2,2\}$ modes decrease as \cite{Baryakhtar:2020gao}
\begin{align}
    \varepsilon_{011}(t>t_2) &\approx \frac{\varepsilon_{011}(t_2)}{\sqrt{1+2\varepsilon_{011}^2(t_2)(t-t_2)/\tau_\mathrm{scalar}}},
    \\
    \varepsilon_{022}(t>t_2) &\approx \frac{\gamma^{022\times\mathrm{BH}}_{011\times011}}{2\gamma^{011\times\infty}_{022\times022}} \varepsilon_{011}(t),
\end{align}
where
\begin{align}
    \tau_\mathrm{scalar} \equiv \frac{4}{3}\frac{\gamma^{011\times\infty}_{022\times022}}{(\gamma^{022\times\mathrm{BH}}_{011\times011})^2}.
\end{align}
Thus, for $t>t_2$, the cumulative GW energy emitted up to time $t$ is
\begin{align}\label{eq:EGW_total_2}
    \begin{split}
        &E_\mathrm{GW}(t>t_2) \approx E_\mathrm{GW}(t_2) + 
        \\
        &\frac{M(t_2)\alpha_2\tau_{\mathrm{scalar}}}{2\varepsilon_{011}(t_1)\tau_{\mathrm{GW},011}}\log\left[1+\frac{2\varepsilon_{011}^2(t_2)(t-t_2)}{\tau_\mathrm{scalar}}\right],
    \end{split}
\end{align}
where $M(t_2)\approx M(t_1)$, $\alpha_2=GM(t_2)\mu_\mathrm{s}$, $\varepsilon_{011}(t_2)$ is obtained from Eq.~\eqref{eq:eps_after_t1}, and $E_\mathrm{GW}(t_2)$ can be calculated from Eq.~\eqref{eq:EGW_total_1}. The mixed annihilation term related to the process in Eq.~\eqref{GW_anni_011022} is retained in the full numerical evolution but is neglected in the analytical evaluation of $E_\mathrm{GW}$. Indeed, during the equilibrium stage described above,
\begin{align}
\frac{\dot E_{011\times022}^{\mathrm{GW}}}
{\dot E_{011\times011}^{\mathrm{GW}}}
\simeq 1.57\times10^{-3}\alpha^2
\frac{\varepsilon_{022}}{\varepsilon_{011}}.
\end{align}
Using the equilibrium relation above, $\varepsilon_{022}/\varepsilon_{011} \lesssim 40\alpha^3$, this ratio is smaller than $3.3\times10^{-5}$ for $\alpha\leq0.22$. Moreover, Fig.~\ref{fig:evo} shows that, for the benchmark evolutions considered here, the $\{0,2,2\}$ occupation remains subdominant in the regime where $\varepsilon_{011}$ is larger and most of the $011\times011$ GW energy is being emitted. We therefore neglect both the direct mixed-channel emission and its backreaction on the analytical $011\times011$ GW energy used in the SGWB calculation.

When $f_\mathrm{s}<f_{\mathrm{s},2}$, the evolution belongs to the large self-coupling regime, as shown in the third row of Fig.~\ref{fig:evo}. The equilibrium of the $\{0,1,1\}$ and $\{0,2,2\}$ modes is established prior to superradiant saturation. As a result, the occupation numbers and then the associated GW emission energy are strongly suppressed. Therefore, we do not consider this regime in the following study of the SGWB. As we show below it is also unlikely that next-generation ground-based GW detectors will be sensitive to decay constants $f_s$ that fall in this regime.

\subsection{Total GW emission energy}
\label{sec:Total_GW}
Based on the results that we just described, we use different prescriptions to compute the total GW emission energy depending on the self-coupling regime. This information is then used in the next section to compute the SGWB emitted by a population of BH-cloud systems.

The required input parameters include the initial BH mass $M_0$, the initial BH spin $a_{*0}$, the BH lifetime $T_\mathrm{BH}$, the scalar mass $\mu_\mathrm{s}$, and the decay constant $f_\mathrm{s}$. First, we compute the critical spin $a_{\mathrm{*c},011}$ using Eq.~\eqref{eq:asc} with the initial parameters. If $a_{*0} < a_{\mathrm{*c},011}$, no cloud is ever formed and the total GW emission energy is $E_\mathrm{GW}=0$. Otherwise, we first compute $f_\mathrm{s,1}$ and $f_\mathrm{s,2}$ using the prescription above and then consider the following procedure:
\begin{itemize}
    \item If $f_\mathrm{s}>f_\mathrm{s,1}$, we compute the time $t_1$. If $T_\mathrm{BH}<t_1$, then $E_\mathrm{GW}=0$; otherwise, $E_\mathrm{GW}$ is computed using Eq.~\eqref{eq:EGW_total_1} with $t=T_\mathrm{BH}$.
    \item If $f_\mathrm{s,2}<f_\mathrm{s}<f_\mathrm{s,1}$,  we set $E_\mathrm{GW} = 0$ for $\alpha>0.22$, to ensure that our two-mode approximation remains valid. As we discuss below, this choice gives a conservative estimate of the overall SGWB. For $\alpha\leq0.22$, we compute the times $t_1$ and $t_2$. If $T_\mathrm{BH}<t_2$, $E_\mathrm{GW}$ is computed following the same procedure as in the $f_\mathrm{s}>f_\mathrm{s,1}$ case; otherwise, $E_\mathrm{GW}$ is computed using Eq.~\eqref{eq:EGW_total_2} with $t=T_\mathrm{BH}$.
    \item If $f_\mathrm{s}<f_\mathrm{s,2}$, then we set $E_\mathrm{GW} = 0$.
\end{itemize}

\section{Stochastic gravitational-wave background}\label{sec:SGWB}

In this section, we first briefly review the SGWB energy density spectrum following Refs.~\cite{Tsukada:2018mbp,Tsukada:2020lgt,Yuan:2021ebu} and present the corresponding results in Sec.~\ref{sec:spectrum}. In Sec.~\ref{sec:SNR}, we calculate the SNR of the SGWB and discuss its detectability as well as projected constraints for different values of the decay constant with LIGO \cite{LIGOScientific:2014pky}, Einstein Telescope (ET) \cite{Punturo:2010zz,ET:2019dnz} and Cosmic Explorer (CE) \cite{Evans:2021gyd,Evans:2023euw}. 

\subsection{Energy density spectrum}
\label{sec:spectrum}

\begin{figure*}
    \centering
    \includegraphics[width=0.48\linewidth]{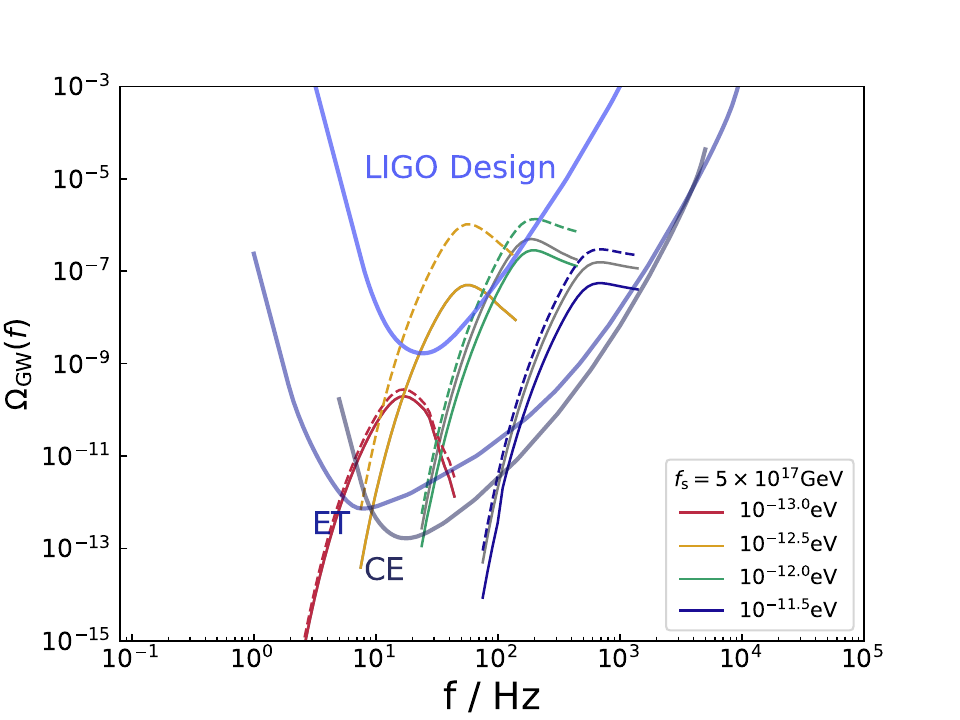}
    \includegraphics[width=0.48\linewidth]{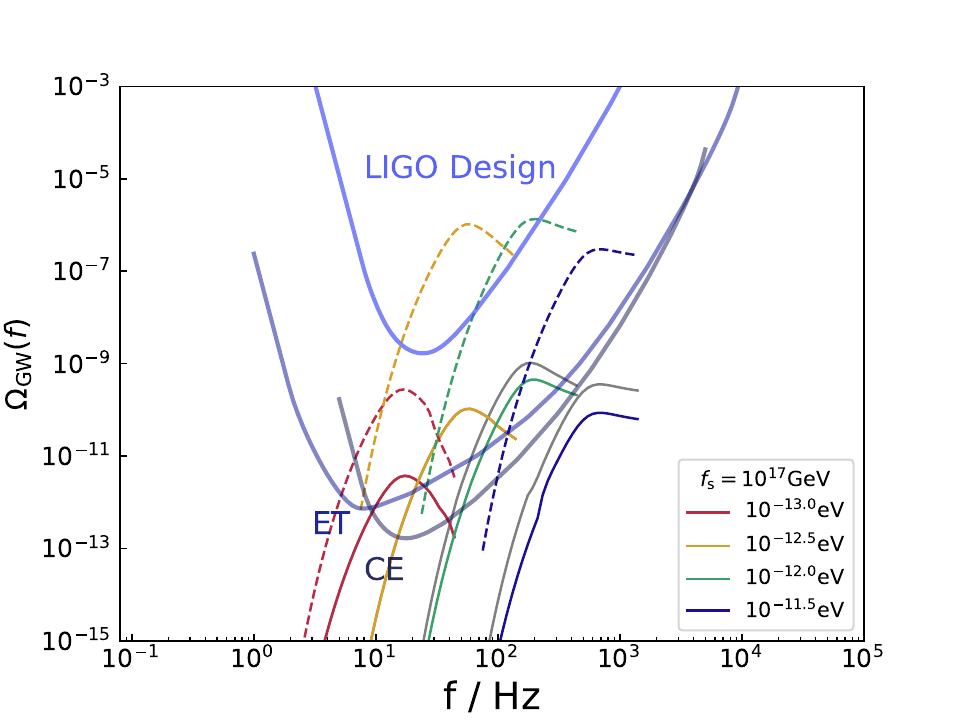}
    \caption{The energy density spectrum of the SGWB sourced by self-interacting scalar clouds. Contributions from both isolated BHs and remnant BHs are included. The thin dashed curves in both panels correspond to a decay constant of $f_{\rm s}=10^{19}\,$GeV. Different colors indicate different scalar masses $\mu_\mathrm{s}$. The thin solid curves correspond to $f_{\rm s}=5\times10^{17}\,$GeV in the left panel and $f_{\rm s}=10^{17}\,$GeV in the right panel. The thin gray solid curves show the spectrum obtained after removing the artificial treatment of setting the GW emission energy to zero for $\alpha>0.22$ in the moderate self-coupling regime. The thick solid lines show the power-law integrated sensitivity curves, assuming a detection threshold of SNR $= 5$ for LIGO, CE, and ET at their design sensitivities. We assume a four-year observation period and two identical, co-aligned, and co-located detectors for CE and ET. For LIGO, we adopt the overlap reduction function computed in Ref.~\cite{Thrane:2013oya}.}
    \label{fig:SGWB}
\end{figure*}

A SGWB can be characterized by its energy density spectrum, 
\begin{align}\label{eq:def_Omega_GW}
    \Omega_\mathrm{GW}(f) \equiv \frac{1}{\rho_\mathrm{c}} \frac{d\rho_\mathrm{GW}}{d\ln(f)},
\end{align}
where $\rho_\mathrm{c}$ is the critical energy density of the Universe, $f$ is the GW frequency observed at the detector, and $\rho_\mathrm{GW}$ is the present-day energy density of GWs. Assuming a homogeneous and isotropic Universe and a random orientation of sources with respect to the observer, the spectrum~\eqref{eq:def_Omega_GW} can be written as \cite{Phinney:2001di,Tsukada:2018mbp,Tsukada:2020lgt}
\begin{align}\label{eq:Omega_GW}
    \Omega_\mathrm{GW}(f) = \frac{f}{\rho_{\mathrm{c}}}\int dz \frac{dt}{dz} \int d\bm{\theta}p(\bm{\theta})R(z;\bm{\theta})\frac{dE_\mathrm{r}}{df_\mathrm{r}}(\bm{\theta}),
\end{align}
where ${dt}/{dz}$ denotes the derivative of the lookback time with respect to the redshift, $p(\bm{\theta})$ represents the joint probability density function of the source parameters $\bm{\theta}$, $R(z;\bm{\theta})$ is the GW event rate per unit comoving volume per unit time in the rest frame of the source, and ${dE_\mathrm{r}}/{df_\mathrm{r}}$ is the energy spectrum of a single GW event in the source rest frame. Since the GW produced by the $011\times011\rightarrow\mathrm{GW}$ process is quasi-monochromatic, the energy spectrum can be approximated by~\cite{Brito:2017wnc}
\begin{align}
    \frac{dE_\mathrm{r}}{df_\mathrm{r}} \approx E_\mathrm{GW} \delta(f(1+z)-f_0),
\end{align}
where $f_0 = \omega_{011}/\pi \approx \mu_{\rm s}/\pi$. Here, we have used the relation between the observed frequency and the frequency in the source rest frame, namely, $f = f_\mathrm{r}/(1+z)$. In addition, in the standard $\Lambda$CDM cosmology, ${dt}/{dz}$ is given by
\begin{align}
    \frac{dt}{dz} = \frac{1}{(1+z)H_0\sqrt{\Omega_\mathrm{m}(1+z)^3+\Omega_{\Lambda}}},
\end{align}
where $\Omega_\mathrm{m}$ and $\Omega_{\Lambda}$ denote the dimensionless density parameters for matter and the cosmological constant, respectively, and $H_0$ is the Hubble constant. In this work, we adopt $\Omega_\mathrm{m}=0.315$, $\Omega_{\Lambda}=0.685$ and $H_0=67.4\,\mathrm{km\,s^{-1}\,Mpc^{-1}}$ \cite{Planck:2018vyg}.

To proceed with the calculation of the SGWB, we apply Eq.~\eqref{eq:Omega_GW} to specific BH population models. Here, we consider two such models: isolated extragalactic BHs formed by core-collapse supernovae and binary BH merger remnants. Due to the non-Gaussian and anisotropic characteristics of the signal from Galactic BHs \cite{Tsukada:2020lgt}, we do not consider the corresponding population (see however Refs.~\cite{Zhu:2020tht,Ellis:2026gwt}).

For isolated extragalactic BHs formed by core-collapse supernovae, Eq.~\eqref{eq:Omega_GW} becomes
\begin{align}
    \Omega^\mathrm{iso}_\mathrm{GW}(f) = \frac{f}{\rho_{\mathrm{c}}}\int dz \frac{dt}{dz} \int dMda_* p(a_*)\frac{d\dot{n}}{dM}\frac{dE_\mathrm{r}}{df_\mathrm{r}},
\end{align}
where $d\dot{n}/dM$ denotes the BH formation rate per comoving volume per BH mass in the source rest frame. We compute this function following Ref.~\cite{Yuan:2021ebu}. Given the limited knowledge of the initial BH spin, we assume a uniform distribution in the range $[0,a_\mathrm{lim}]$, where $a_\mathrm{lim} \simeq 0.998$ is the Thorne limit of the BH spin \cite{Thorne:1974ve}.

For binary BH merger remnants, Eq.~\eqref{eq:Omega_GW} can be written as
\begin{align}
    \Omega^\mathrm{rem}_\mathrm{GW}(f) = \frac{f}{\rho_{\mathrm{c}}}\int dz \frac{dt}{dz} \int dM_1dM_2 \mathcal{R}(z,M_1,M_2) \frac{dE_\mathrm{r}}{df_\mathrm{r}},
\end{align}
where $\mathcal{R}(z,M_1,M_2)$ denotes the merger rate density at redshift $z$ for progenitor masses $M_1$ and $M_2$. We compute this rate following Ref.~\cite{Yuan:2021ebu}, except that we use an updated local merger rate, $19\,\mathrm{yr}^{-1}\,\mathrm{Gpc}^{-3}$ \cite{LIGOScientific:2025pvj}. For a given pair of progenitor masses, Numerical Relativity simulations provide the initial mass $M_{0}$ and the initial spin $a_{*0}$ of the remnant BH formed after the binary merger \cite{Berti:2007fi,Scheel:2008rj,Barausse:2009uz,Yuan:2021ebu},
\begin{align}
    \begin{split}
        M_0 &= M_1 + M_2 - (M_1+M_2)\Bigg[\left(1-\sqrt{\frac{8}{9}}\right)\nu
        \\
        &\hspace{0.4cm} + 4\nu^2\left(0.19308+\sqrt{\frac{8}{9}}-1\right)\Bigg],
    \end{split}
    \\
    a_{*0} &= \nu (2\sqrt{3}-3.5171\nu + 2.5763\nu^2),
\end{align}
where $\nu\equiv M_1M_2/(M_1+M_2)^2$, and the effects of the progenitor spins are neglected.

The observed SGWB is expected to be a superposition of contributions from these two BH population models, and therefore the total energy density spectrum is given by
\begin{align}
    \Omega_\mathrm{GW}(f) = \Omega_\mathrm{GW}^\mathrm{iso}(f) + \Omega_\mathrm{GW}^\mathrm{rem}(f),
\end{align}
The resulting spectrum is shown in Fig.~\ref{fig:SGWB}. The thin dashed curves in both panels correspond to a decay constant of $f_{\rm s}=10^{19}\,$GeV, which effectively corresponds to the scenario without self-interaction. Different colors represent different scalar masses $\mu_\mathrm{s}$. All four scalar-mass cases considered are potentially observable by ET and CE. In addition, LIGO is sensitive to the cases with $\mu_\mathrm{s}=10^{-12.5}\,$eV and $\mu_\mathrm{s}=10^{-12}\,$eV. However, the null results from previous SGWB searches reported in Refs.~\cite{Tsukada:2018mbp,Yuan:2022bem} indicate that these two cases are ruled out in the absence of self-interactions.

The results also show that a smaller decay constant can significantly relax LVK's constraints on the scalar field mass. The thin solid curves correspond to a decay constant of $f_{\rm s}=5\times10^{17}\,$GeV in the left panel and $f_{\rm s}=10^{17}\,$GeV in the right panel. Since the stage during which GW emission dominates the evolution of the system is shortened by moderate self-interactions, the GW energy density spectrum decreases as the decay constant becomes smaller. In particular, for the scenario with $f_{\rm s}=10^{17}\,$GeV, all scalar-mass cases considered in Fig.~\ref{fig:SGWB} lie below the design sensitivity of LIGO, while still remaining above the sensitivities of ET and CE except for the case $\mu_\mathrm{s}=10^{-11.5}\,$eV. This demonstrates that the constraints from LVK on scalar fields can be substantially relaxed once self-interactions are taken into account.

Finally, to assess how our treatment of systems with $\alpha>0.22$ in the moderate self-coupling regime affects our results, we perform a simple check. Although a more detailed analysis beyond the two-mode approximation would be required for a complete treatment, we can extrapolate our moderate self-coupling prescription to $\alpha>0.22$ by removing the \emph{ad hoc} assumption that the GW emission energy vanishes in this region. The resulting spectra are shown as thin gray curves in Fig.~\ref{fig:SGWB}. For $\mu_\mathrm{s}=10^{-13}\,$eV and $\mu_\mathrm{s}=10^{-12.5}\,$eV, the thin colored and thin gray curves coincide, because the background in these cases is dominated by systems with $\alpha\lesssim 0.22$. As the scalar mass increases, and particularly for $\mu_\mathrm{s}\gtrsim 10^{-12}\,$ eV, the contribution from the $\alpha>0.22$ region becomes more important. Even so, it only slightly changes the overall amplitude of the background, and only for a narrow range of detectable scalar masses, so our main conclusions are mostly unaffected. Our results, which neglect the contribution from $\alpha>0.22$, should therefore be regarded as a conservative estimate.

\subsection{Detectability and projected constraints}
\label{sec:SNR}

\begin{figure}[!h]
    \centering
    \includegraphics[width=0.99\linewidth]{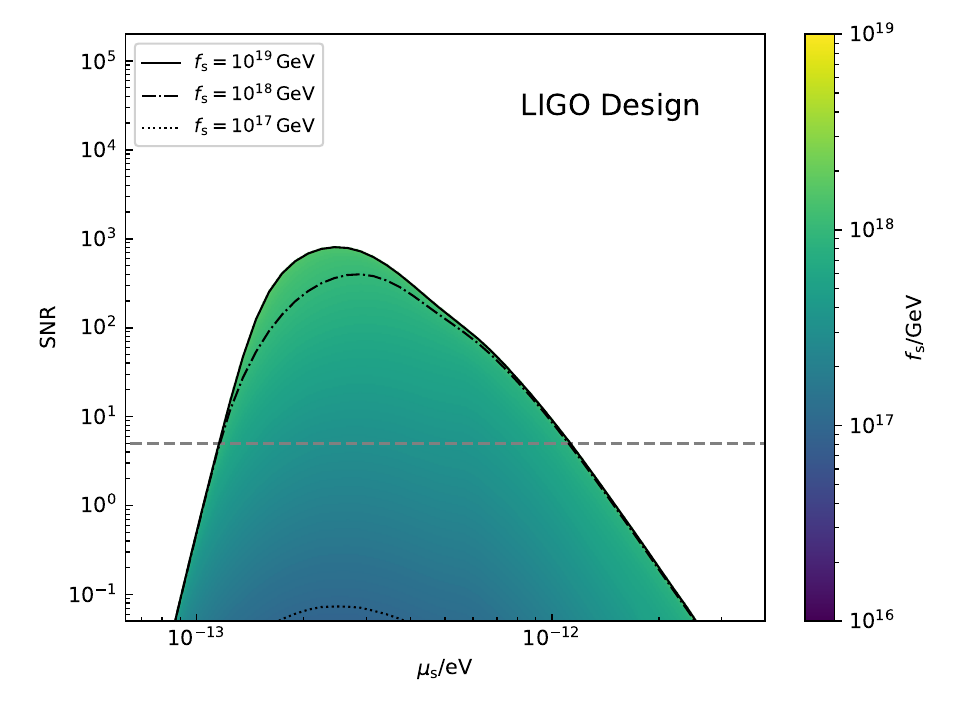}
    \includegraphics[width=0.99\linewidth]{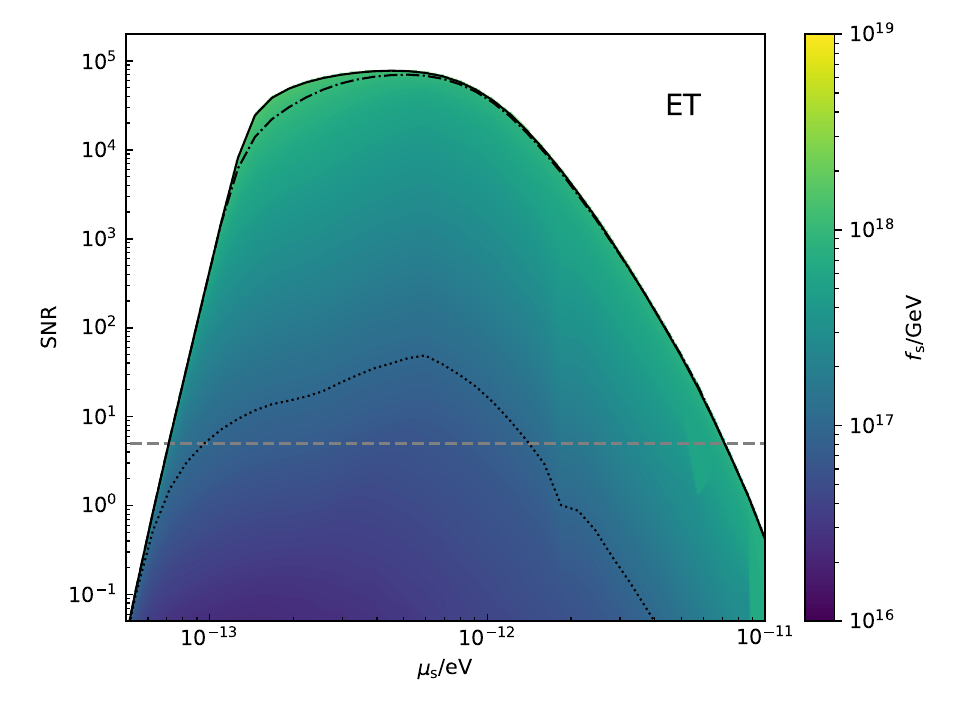}
    \includegraphics[width=0.99\linewidth]{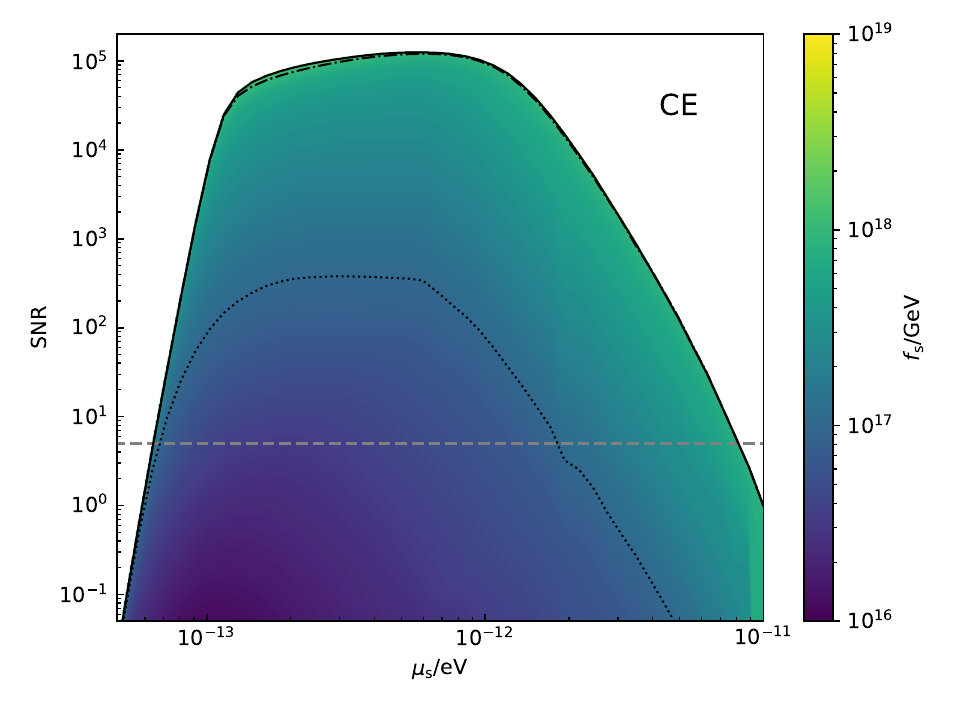}
    \caption{SNR of the SGWB sourced by a population of self-interacting scalar clouds as a function of the scalar mass $\mu_\mathrm{s}$ and decay constant $f_\mathrm{s}$ for LIGO at design sensitivity (top), ET (middle), and CE (bottom). Contributions from both isolated BHs and binary BH merger remnants are included. Different colors denote different values of the decay constant. The black solid, dash-dotted, and dotted curves correspond to $f_\mathrm{s}=10^{19}$, $10^{18}$, and $10^{17}\,\mathrm{GeV}$, respectively. The gray dashed line denotes SNR $=5$. For all detectors, we assume an observation time of $T_{\rm obs}=4$ yr.}
    \label{fig:SNR}
\end{figure}

To further investigate the effect of self-interactions on the detectability of the SGWB and provide a projection of possible constraints with LIGO, CE and ET, we compute the SNR of the signal, which quantifies the significance of the SGWB signal relative to (non-astrophysical) noise fluctuations. For an arbitrarily large SGWB, the SNR is given by \cite{Allen:1997ad,Yuan:2021ebu}
\begin{align} 
    \begin{split}
        \mathrm{SNR}^2 &= 2T_\mathrm{obs} \times 
        \\
        \int_0^{\infty} df &\frac{\Gamma_{IJ}(f)^2 S_h(f)^2}{\left[\frac{1}{25}+\Gamma_{IJ}(f)^2\right]S_h(f)^2+P_n(f)^2+\frac{2}{5}S_h(f)P_n(f)},
    \end{split}
\end{align}
where $T_\mathrm{obs}$ is the observation time, $\Gamma_{IJ}(f)$ is the (unnormalized) overlap reduction function between detectors $I$ and $J$, and $P_n(f)$ is the one-sided noise power spectral density. The GW strain power spectral density is related to the energy density spectrum by
\begin{align} 
        S_h(f)=\frac{3H_0^2}{2\pi^2 f^3}\Omega_\mathrm{GW}(f).
\end{align}
We take the observation time to be $T_\mathrm{obs} =4$ yr. Unless otherwise stated, we assume two identical interferometers with opening angle $\delta=90^{\circ}$. For CE and ET, we assume they are two co-located and co-aligned detectors, and this implies $\Gamma_{IJ}(0)=1/5$.

Figure~\ref{fig:SNR} shows the SNR as a function of the scalar mass and decay constant for LIGO at design sensitivity (top), ET (middle), and CE (bottom). Different colors denote different values of the decay constant. We adopt SNR $=5$ as the detectability threshold, indicated by a gray dashed line in each panel. Across all three detectors, there is a common trend: as the decay constant decreases, the SNR decreases and may fall below the threshold. For LIGO at design sensitivity, the detector is sensitive to scalar masses in the range $\sim[10^{-12.9},10^{-11.9}]$ eV when $f_\mathrm{s}=10^{19}\,$GeV. For the smaller decay constant $f_\mathrm{s}=3.06\times10^{17}\,$GeV, the SNR is reduced, with a maximum value of $\sim5$ within the scalar-mass range shown. In other words, for $f_\mathrm{s}$ below this value, the SGWB falls below the detectability threshold regardless of the boson mass. As discussed above, this indicates a smaller decay constant can significantly relax LVK's constraints on scalars.

On the other hand, the SGWB sourced by a moderately self-interacting scalar cloud remains potentially detectable by ET and CE. ET would be sensitive to scalar masses in the range $\sim[10^{-13.2},10^{-11.1}]$ eV when $f_\mathrm{s}=10^{19}\,$GeV, and $\sim[10^{-13.0},10^{-11.8}]$ eV when $f_\mathrm{s}=10^{17}\,$ GeV. For CE, the ranges change to $\sim[10^{-13.2},10^{-11.1}]$ eV when $f_\mathrm{s}=10^{19}\,$GeV, and $\sim[10^{-13.2},10^{-11.7}]$ eV when $f_\mathrm{s}=10^{17}\,$GeV. The minimum decay constants that still yield a detectable SGWB for ET and CE are $f_\mathrm{s}=5.90\times10^{16}\,$GeV and $3.20\times10^{16}\,$GeV, respectively, which is roughly one order of magnitude better than LIGO.

Finally, we note that relativistic calculations indicate that additional modes can become relevant already at $\alpha\sim0.15$~\cite{Witte:2024drg}. To assess the sensitivity of our predictions to this limitation, we repeat the calculation with the coupling cutoff in the moderate self-coupling regime lowered from $0.22$ to $0.15$, keeping all other assumptions unchanged. For $f_\mathrm{s}=10^{17}\,\mathrm{GeV}$, the effect is negligible at $\mu_\mathrm{s}\lesssim 10^{-12.4}\,\mathrm{eV}$, but becomes increasingly important at higher scalar masses. In particular, at $\mu_\mathrm{s}=10^{-12.5}\,\mathrm{eV}$, the ET and CE SNRs decrease by less than $0.1\%$, whereas at $\mu_\mathrm{s}=10^{-12}\,\mathrm{eV}$ they decrease by approximately $75.2\%$ and $75.9\%$, respectively. Thus, the potentially detectable signal at lower scalar masses survives this check, while the higher-mass reach is sensitive to the adopted cutoff. A more reliable prediction in this higher-mass region therefore requires relativistic calculations and the coupled evolution of additional modes.

\section{Summary and discussion}
\label{sec:Summary}

In this work, we investigated the SGWB produced by BH-cloud systems, in which a rotating BH is surrounded by a cloud of an ultralight scalar field. Unlike previous studies \cite{Brito:2017wnc,Brito:2017zvb,Tsukada:2018mbp,Yuan:2021ebu,Yuan:2022bem}, where self-interactions were neglected, we quantified the impact of scalar self-interactions. Self-interactions modify the evolution of the cloud and, if sufficiently strong, can limit its growth, thereby suppressing the GW emission from the annihilation of scalar quanta~\cite{Baryakhtar:2020gao}. By computing the SGWB energy density spectrum and corresponding SNR at different detectors, we found that self-interactions can relax LVK's constraints on scalar fields derived from the null searches in Refs.~\cite{Tsukada:2018mbp,Yuan:2022bem}. In particular, for decay constants $f_\mathrm{s}\lesssim 3\times10^{17}\,$GeV no constraints can be imposed with LIGO at design sensitivity.

Moreover, we showed that for future detectors such as ET and CE, the SGWB sourced by a moderately self-interacting scalar cloud remains potentially detectable. For example, taking $f_\mathrm{s} = 10^{17}$ GeV, ET could detect or constrain scalar fields with masses in the range $\sim[10^{-13.0},10^{-11.8}]$ eV, while CE would be sensitive to the range $\sim[10^{-13.2},10^{-11.7}]$ eV. Considering all possible scalar masses and the detectors we considered, the minimum decay constant that could still lead to a potentially detectable SGWB is $f_\mathrm{s}\sim 3\times10^{16}\,$GeV.

There remains room for improvement in this work. We employed a two-mode approximation in the small-$\alpha$ approximation, which is valid for $\alpha \lesssim 0.22$ and for evolution that begins with the $\{0,1,1\}$ mode growing. For $\alpha \gtrsim 0.22$ in the moderate self-coupling regime, we set the GW emission energy to zero, and we applied the same setting in the large self-coupling regime. Going beyond the two-mode approximation would be required to handle the regime $\alpha \gtrsim 0.22$. In addition, we also neglected GW emission from high-$m$ modes. These artificial settings suppress the predicted SGWB for a small subset of potentially detectable scalar masses.
The detectable parameter space for next-generation detectors derived here should therefore be regarded as conservative. Moreover, when relativistic corrections are included, the two-mode approximation starts instead to break down at $\alpha \sim 0.15$ \cite{Witte:2024drg}, suggesting that relativistic corrections might also be important for part of the parameter space. Finally, low-frequency transition signals from the $022\rightarrow011\times\mathrm{GW}$ may also be detectable, for example with space-based GW detectors. We leave these improvements to future work.

\section*{Acknowledgments}
We acknowledge the use of the \texttt{GWSC.jl} package \footnote{\url{https://github.com/zuchengchen/GWSC.jl}} in calculating the sensitivity curves.
Y.G. is deeply grateful to Vitor Cardoso and R.B. for their warm hospitality during his stay at CENTRA/IST. He acknowledges financial support from the National Natural Science Foundation of China (Grants Nos. 124B2098, 12447105, 12075136) and from the Natural Science Foundation of Shandong Province (Grant No. ZR2020MA094). 
R.B. acknowledges financial support provided by FCT – Fundação para a Ciência e a Tecnologia, I.P., through the ERC-Portugal program Project ``GravNewFields''.
C.Y. acknowledges the financial support provided under the European Union’s H2020 ERC Advanced Grant “Black holes: gravitational engines of discovery” grant agreement no. Gravitas–101052587. Views and opinions expressed are however those of the author only and do not necessarily reflect those of the European Union or the European Research Council. Neither the European Union nor the granting authority can be held responsible for them. He also acknowledges support from the Villum Investigator program supported by the VILLUM Foundation (grant no. VIL37766) and the DNRF Chair program (grant no. DNRF162) by the Danish National Research Foundation.
We also thank the Fundação para a Ciência e Tecnologia (FCT), Portugal, for the financial support to the Center for Astrophysics and Gravitation (CENTRA/IST/ULisboa) through grant No.~\href{https://doi.org/10.54499/UID/PRR/00099/2025}{UID/PRR/00099/2025} and grant No.~\href{https://doi.org/10.54499/UID/00099/2025}{UID/00099/2025}.

\bibliography{refs}

@article{LIGOScientific:2016aoc,
    author = "Abbott, B. P. and others",
    collaboration = "LIGO Scientific, Virgo",
    title = "{Observation of Gravitational Waves from a Binary Black Hole Merger}",
    eprint = "1602.03837",
    archivePrefix = "arXiv",
    primaryClass = "gr-qc",
    reportNumber = "LIGO-P150914",
    doi = "10.1103/PhysRevLett.116.061102",
    journal = "Phys. Rev. Lett.",
    volume = "116",
    number = "6",
    pages = "061102",
    year = "2016"
}

@article{LIGOScientific:2018mvr,
    author = "Abbott, B. P. and others",
    collaboration = "LIGO Scientific, Virgo",
    title = "{GWTC-1: A Gravitational-Wave Transient Catalog of Compact Binary Mergers Observed by LIGO and Virgo during the First and Second Observing Runs}",
    eprint = "1811.12907",
    archivePrefix = "arXiv",
    primaryClass = "astro-ph.HE",
    reportNumber = "LIGO-P1800307",
    doi = "10.1103/PhysRevX.9.031040",
    journal = "Phys. Rev. X",
    volume = "9",
    number = "3",
    pages = "031040",
    year = "2019"
}

@article{LIGOScientific:2020ibl,
    author = "Abbott, R. and others",
    collaboration = "LIGO Scientific, Virgo",
    title = "{GWTC-2: Compact Binary Coalescences Observed by LIGO and Virgo During the First Half of the Third Observing Run}",
    eprint = "2010.14527",
    archivePrefix = "arXiv",
    primaryClass = "gr-qc",
    reportNumber = "P2000061",
    doi = "10.1103/PhysRevX.11.021053",
    journal = "Phys. Rev. X",
    volume = "11",
    pages = "021053",
    year = "2021"
}

@article{LIGOScientific:2021usb,
    author = "Abbott, R. and others",
    collaboration = "LIGO Scientific, VIRGO",
    title = "{GWTC-2.1: Deep extended catalog of compact binary coalescences observed by LIGO and Virgo during the first half of the third observing run}",
    eprint = "2108.01045",
    archivePrefix = "arXiv",
    primaryClass = "gr-qc",
    reportNumber = "LIGO-P2100063",
    doi = "10.1103/PhysRevD.109.022001",
    journal = "Phys. Rev. D",
    volume = "109",
    number = "2",
    pages = "022001",
    year = "2024"
}

@article{KAGRA:2021vkt,
    author = "Abbott, R. and others",
    collaboration = "KAGRA, VIRGO, LIGO Scientific",
    title = "{GWTC-3: Compact Binary Coalescences Observed by LIGO and Virgo during the Second Part of the Third Observing Run}",
    eprint = "2111.03606",
    archivePrefix = "arXiv",
    primaryClass = "gr-qc",
    reportNumber = "LIGO-P2000318",
    doi = "10.1103/PhysRevX.13.041039",
    journal = "Phys. Rev. X",
    volume = "13",
    number = "4",
    pages = "041039",
    year = "2023"
}

@article{LIGOScientific:2025slb,
    author = "Abac, A. G. and others",
    collaboration = "LIGO Scientific, VIRGO, KAGRA",
    title = "{GWTC-4.0: Updating the Gravitational-Wave Transient Catalog with Observations from the First Part of the Fourth LIGO-Virgo-KAGRA Observing Run}",
    eprint = "2508.18082",
    archivePrefix = "arXiv",
    primaryClass = "gr-qc",
    reportNumber = "LIGO-P2400386",
    doi = "10.3847/2041-8213/ae2c74",
    journal = "Astrophys. J. Lett.",
    volume = "1004",
    number = "2",
    pages = "L22",
    year = "2026"
}

@article{LIGOScientific:2026wfs,
    author = "Abac, A. G. and others",
    collaboration = "LIGO Scientific, VIRGO, KAGRA",
    title = "{GWTC-5.0: Observations from the Second Part of the Fourth LIGO-Virgo-KAGRA Observing Run and Updates to the Gravitational-Wave Transient Catalog}",
    eprint = "2605.27225",
    archivePrefix = "arXiv",
    primaryClass = "gr-qc",
    reportNumber = "LIGO-P2600152",
    month = "5",
    year = "2026"
}

@article{LIGOScientific:2026mjf,
    author = "Abac, A. G. and others",
    collaboration = "LIGO Scientific, VIRGO, KAGRA",
    title = "{Updated Upper Limits on the Isotropic Gravitational-Wave Background from LIGO, Virgo, and KAGRA Data through April 2025}",
    eprint = "2608.23477",
    archivePrefix = "arXiv",
    primaryClass = "gr-qc",
    reportNumber = "LIGO P2600217-v10",
    month = "8",
    year = "2026"
}

@article{Christensen:2018iqi,
    author = "Christensen, Nelson",
    title = "{Stochastic Gravitational Wave Backgrounds}",
    eprint = "1811.08797",
    archivePrefix = "arXiv",
    primaryClass = "gr-qc",
    doi = "10.1088/1361-6633/aae6b5",
    journal = "Rept. Prog. Phys.",
    volume = "82",
    number = "1",
    pages = "016903",
    year = "2019"
}

@article{East:2017ovw,
    author = "East, William E. and Pretorius, Frans",
    title = "{Superradiant Instability and Backreaction of Massive Vector Fields around Kerr Black Holes}",
    eprint = "1704.04791",
    archivePrefix = "arXiv",
    primaryClass = "gr-qc",
    doi = "10.1103/PhysRevLett.119.041101",
    journal = "Phys. Rev. Lett.",
    volume = "119",
    number = "4",
    pages = "041101",
    year = "2017"
}

@article{Herdeiro:2017phl,
    author = "Herdeiro, Carlos A. R. and Radu, Eugen",
    title = "{Dynamical Formation of Kerr Black Holes with Synchronized Hair: An Analytic Model}",
    eprint = "1706.06597",
    archivePrefix = "arXiv",
    primaryClass = "gr-qc",
    doi = "10.1103/PhysRevLett.119.261101",
    journal = "Phys. Rev. Lett.",
    volume = "119",
    number = "26",
    pages = "261101",
    year = "2017"
}

@article{Guo:2025dkx,
    author = "Guo, Yin-Da and Bao, Shou-Shan and Li, Tianjun and Zhang, Hong",
    title = "{Effect of accretion on scalar superradiant instability}",
    eprint = "2501.09280",
    archivePrefix = "arXiv",
    primaryClass = "gr-qc",
    doi = "10.1088/1475-7516/2025/09/066",
    journal = "JCAP",
    volume = "09",
    pages = "066",
    year = "2025"
}

@article{Zeldovich:1971ffh,
    author = "Zeldovich, Yakov Borisovich",
    title = "{Generation of Waves by a Rotating Body}",
    journal = "Soviet Journal of Experimental and Theoretical Physics Letters",
    volume = "14",
    pages = "180",
    year = "1971"
}

@article{Zeldovich:1972zqp,
    author = "Zeldovich, Yakov Borisovich",
    title = "{Amplification of Cylindrical Electromagnetic Waves Reflected from a Rotating Body}",
    journal = "Soviet Journal of Experimental and Theoretical Physics",
    volume = "35",
    pages = "1085",
    year = "1972"
}

@article{Press:1972zz,
    author = "Press, William H. and Teukolsky, Saul A.",
    title = "{Floating Orbits, Superradiant Scattering and the Black-hole Bomb}",
    doi = "10.1038/238211a0",
    journal = "Nature",
    volume = "238",
    pages = "211--212",
    year = "1972"
}

@article{Detweiler:1980uk,
    author = "Detweiler, Steven L.",
    title = "{KLEIN-GORDON EQUATION AND ROTATING BLACK HOLES}",
    doi = "10.1103/PhysRevD.22.2323",
    journal = "Phys. Rev. D",
    volume = "22",
    pages = "2323--2326",
    year = "1980"
}

@article{Dolan:2007mj,
    author = "Dolan, Sam R.",
    title = "{Instability of the massive Klein-Gordon field on the Kerr spacetime}",
    eprint = "0705.2880",
    archivePrefix = "arXiv",
    primaryClass = "gr-qc",
    doi = "10.1103/PhysRevD.76.084001",
    journal = "Phys. Rev. D",
    volume = "76",
    pages = "084001",
    year = "2007"
}

@article{Brito:2014wla,
    author = "Brito, Richard and Cardoso, Vitor and Pani, Paolo",
    title = "{Black holes as particle detectors: evolution of superradiant instabilities}",
    eprint = "1411.0686",
    archivePrefix = "arXiv",
    primaryClass = "gr-qc",
    doi = "10.1088/0264-9381/32/13/134001",
    journal = "Class. Quant. Grav.",
    volume = "32",
    number = "13",
    pages = "134001",
    year = "2015"
}

@article{Brito:2015oca,
    author = "Brito, Richard and Cardoso, Vitor and Pani, Paolo",
    title = "{Superradiance}: {New Frontiers in Black Hole
Physics}",
    eprint = "1501.06570",
    archivePrefix = "arXiv",
    primaryClass = "gr-qc",
    doi = "10.1007/978-3-319-19000-6",
    journal = "Lect. Notes Phys.",
    volume = "906",
    pages = "pp.1--237",
    year = "2015"
}

@article{Arvanitaki:2010sy,
    author = "Arvanitaki, Asimina and Dubovsky, Sergei",
    title = "{Exploring the String Axiverse with Precision Black Hole Physics}",
    eprint = "1004.3558",
    archivePrefix = "arXiv",
    primaryClass = "hep-th",
    doi = "10.1103/PhysRevD.83.044026",
    journal = "Phys. Rev. D",
    volume = "83",
    pages = "044026",
    year = "2011"
}

@article{Yoshino:2013ofa,
    author = "Yoshino, Hirotaka and Kodama, Hideo",
    title = "{Gravitational radiation from an axion cloud around a black hole: Superradiant phase}",
    eprint = "1312.2326",
    archivePrefix = "arXiv",
    primaryClass = "gr-qc",
    reportNumber = "KEK-TH-1694",
    doi = "10.1093/ptep/ptu029",
    journal = "PTEP",
    volume = "2014",
    pages = "043E02",
    year = "2014"
}

@article{Yoshino:2014wwa,
    author = "Yoshino, Hirotaka and Kodama, Hideo",
    title = "{Probing the string axiverse by gravitational waves from Cygnus X-1}",
    eprint = "1407.2030",
    archivePrefix = "arXiv",
    primaryClass = "gr-qc",
    reportNumber = "KEK-TH-1751, KEK-COSMO-151",
    doi = "10.1093/ptep/ptv067",
    journal = "PTEP",
    volume = "2015",
    number = "6",
    pages = "061E01",
    year = "2015"
}

@article{Arvanitaki:2014wva,
    author = "Arvanitaki, Asimina and Baryakhtar, Masha and Huang, Xinlu",
    title = "{Discovering the QCD Axion with Black Holes and Gravitational Waves}",
    eprint = "1411.2263",
    archivePrefix = "arXiv",
    primaryClass = "hep-ph",
    doi = "10.1103/PhysRevD.91.084011",
    journal = "Phys. Rev. D",
    volume = "91",
    number = "8",
    pages = "084011",
    year = "2015"
}

@article{Arvanitaki:2016qwi,
    author = "Arvanitaki, Asimina and Baryakhtar, Masha and Dimopoulos, Savas and Dubovsky, Sergei and Lasenby, Robert",
    title = "{Black Hole Mergers and the QCD Axion at Advanced LIGO}",
    eprint = "1604.03958",
    archivePrefix = "arXiv",
    primaryClass = "hep-ph",
    doi = "10.1103/PhysRevD.95.043001",
    journal = "Phys. Rev. D",
    volume = "95",
    number = "4",
    pages = "043001",
    year = "2017"
}

@article{Brito:2017wnc,
    author = "Brito, Richard and Ghosh, Shrobana and Barausse, Enrico and Berti, Emanuele and Cardoso, Vitor and Dvorkin, Irina and Klein, Antoine and Pani, Paolo",
    title = "{Stochastic and resolvable gravitational waves from ultralight bosons}",
    eprint = "1706.05097",
    archivePrefix = "arXiv",
    primaryClass = "gr-qc",
    doi = "10.1103/PhysRevLett.119.131101",
    journal = "Phys. Rev. Lett.",
    volume = "119",
    number = "13",
    pages = "131101",
    year = "2017"
}

@article{Brito:2017zvb,
    author = "Brito, Richard and Ghosh, Shrobana and Barausse, Enrico and Berti, Emanuele and Cardoso, Vitor and Dvorkin, Irina and Klein, Antoine and Pani, Paolo",
    title = "{Gravitational wave searches for ultralight bosons with LIGO and LISA}",
    eprint = "1706.06311",
    archivePrefix = "arXiv",
    primaryClass = "gr-qc",
    doi = "10.1103/PhysRevD.96.064050",
    journal = "Phys. Rev. D",
    volume = "96",
    number = "6",
    pages = "064050",
    year = "2017"
}

@article{Baryakhtar:2017ngi,
    author = "Baryakhtar, Masha and Lasenby, Robert and Teo, Mae",
    title = "{Black Hole Superradiance Signatures of Ultralight Vectors}",
    eprint = "1704.05081",
    archivePrefix = "arXiv",
    primaryClass = "hep-ph",
    doi = "10.1103/PhysRevD.96.035019",
    journal = "Phys. Rev. D",
    volume = "96",
    number = "3",
    pages = "035019",
    year = "2017"
}

@article{Isi:2018pzk,
    author = "Isi, Maximiliano and Sun, Ling and Brito, Richard and Melatos, Andrew",
    title = "{Directed searches for gravitational waves from ultralight bosons}",
    eprint = "1810.03812",
    archivePrefix = "arXiv",
    primaryClass = "gr-qc",
    reportNumber = "LIGO-P1800270",
    doi = "10.1103/PhysRevD.99.084042",
    journal = "Phys. Rev. D",
    volume = "99",
    number = "8",
    pages = "084042",
    year = "2019",
    note = "[Erratum: Phys.Rev.D 102, 049901 (2020)]"
}

@article{Palomba:2019vxe,
    author = "Palomba, Cristiano and others",
    title = "{Direct constraints on ultra-light boson mass from searches for continuous gravitational waves}",
    eprint = "1909.08854",
    archivePrefix = "arXiv",
    primaryClass = "astro-ph.HE",
    doi = "10.1103/PhysRevLett.123.171101",
    journal = "Phys. Rev. Lett.",
    volume = "123",
    pages = "171101",
    year = "2019"
}

@article{Siemonsen:2019ebd,
    author = "Siemonsen, Nils and East, William E.",
    title = "{Gravitational wave signatures of ultralight vector bosons from black hole superradiance}",
    eprint = "1910.09476",
    archivePrefix = "arXiv",
    primaryClass = "gr-qc",
    doi = "10.1103/PhysRevD.101.024019",
    journal = "Phys. Rev. D",
    volume = "101",
    number = "2",
    pages = "024019",
    year = "2020"
}

@article{Sun:2019mqb,
    author = "Sun, Ling and Brito, Richard and Isi, Maximiliano",
    title = "{Search for ultralight bosons in Cygnus X-1 with Advanced LIGO}",
    eprint = "1909.11267",
    archivePrefix = "arXiv",
    primaryClass = "gr-qc",
    doi = "10.1103/PhysRevD.101.063020",
    journal = "Phys. Rev. D",
    volume = "101",
    number = "6",
    pages = "063020",
    year = "2020",
    note = "[Erratum: Phys.Rev.D 102, 089902 (2020)]"
}

@article{Brito:2020lup,
    author = "Brito, Richard and Grillo, Sara and Pani, Paolo",
    title = "{Black Hole Superradiant Instability from Ultralight Spin-2 Fields}",
    eprint = "2002.04055",
    archivePrefix = "arXiv",
    primaryClass = "gr-qc",
    doi = "10.1103/PhysRevLett.124.211101",
    journal = "Phys. Rev. Lett.",
    volume = "124",
    number = "21",
    pages = "211101",
    year = "2020"
}

@article{Baryakhtar:2020gao,
    author = "Baryakhtar, Masha and Galanis, Marios and Lasenby, Robert and Simon, Olivier",
    title = "{Black hole superradiance of self-interacting scalar fields}",
    eprint = "2011.11646",
    archivePrefix = "arXiv",
    primaryClass = "hep-ph",
    doi = "10.1103/PhysRevD.103.095019",
    journal = "Phys. Rev. D",
    volume = "103",
    number = "9",
    pages = "095019",
    year = "2021"
}

@article{Zhu:2020tht,
    author = "Zhu, Sylvia J. and Baryakhtar, Masha and Papa, Maria Alessandra and Tsuna, Daichi and Kawanaka, Norita and Eggenstein, Heinz-Bernd",
    title = "{Characterizing the continuous gravitational-wave signal from boson clouds around Galactic isolated black holes}",
    eprint = "2003.03359",
    archivePrefix = "arXiv",
    primaryClass = "gr-qc",
    doi = "10.1103/PhysRevD.102.063020",
    journal = "Phys. Rev. D",
    volume = "102",
    number = "6",
    pages = "063020",
    year = "2020"
}

@article{LIGOScientific:2021rnv,
    author = "Abbott, R. and others",
    collaboration = "LIGO Scientific, Virgo, KAGRA",
    title = "{All-sky search for gravitational wave emission from scalar boson clouds around spinning black holes in LIGO O3 data}",
    eprint = "2111.15507",
    archivePrefix = "arXiv",
    primaryClass = "astro-ph.HE",
    reportNumber = "P2100343",
    doi = "10.1103/PhysRevD.105.102001",
    journal = "Phys. Rev. D",
    volume = "105",
    number = "10",
    pages = "102001",
    year = "2022"
}

@article{KAGRA:2022osp,
    author = "Abbott, R. and others",
    collaboration = "KAGRA, LIGO Scientific, VIRGO",
    title = "{Search for continuous gravitational wave emission from the Milky~Way center in O3 LIGO-Virgo data}",
    eprint = "2204.04523",
    archivePrefix = "arXiv",
    primaryClass = "astro-ph.HE",
    doi = "10.1103/PhysRevD.106.042003",
    journal = "Phys. Rev. D",
    volume = "106",
    number = "4",
    pages = "042003",
    year = "2022"
}

@article{Collaviti:2024mvh,
    author = "Collaviti, Spencer and Sun, Ling and Galanis, Marios and Baryakhtar, Masha",
    title = "{Observational prospects of self-interacting scalar superradiance with next-generation gravitational-wave detectors}",
    eprint = "2407.04304",
    archivePrefix = "arXiv",
    primaryClass = "gr-qc",
    doi = "10.1088/1361-6382/ad96ff",
    journal = "Class. Quant. Grav.",
    volume = "42",
    number = "2",
    pages = "025006",
    year = "2025"
}

@article{Jones:2024fpg,
    author = "Jones, Dana and Siemonsen, Nils and Sun, Ling and East, William E. and Miller, Andrew L. and Wette, Karl and Piccinni, Ornella J.",
    title = "{Methodology for constraining ultralight vector bosons with gravitational wave searches targeting merger remnant black holes}",
    eprint = "2412.00320",
    archivePrefix = "arXiv",
    primaryClass = "gr-qc",
    doi = "10.1103/PhysRevD.111.063028",
    journal = "Phys. Rev. D",
    volume = "111",
    number = "6",
    pages = "063028",
    year = "2025"
}

@article{Omiya:2024xlz,
    author = "Omiya, Hidetoshi and Takahashi, Takuya and Tanaka, Takahiro and Yoshino, Hirotaka",
    title = "{Deci-Hz gravitational waves from the self-interacting axion cloud around a rotating stellar mass black hole}",
    eprint = "2404.16265",
    archivePrefix = "arXiv",
    primaryClass = "gr-qc",
    doi = "10.1103/PhysRevD.110.044002",
    journal = "Phys. Rev. D",
    volume = "110",
    number = "4",
    pages = "044002",
    year = "2024"
}

@article{Guo:2024dqd,
    author = "Guo, Yin-Da and Jia, Nayun and Bao, Shou-Shan and Zhang, Hong and Zhang, Xin",
    title = "{Evolution and detection of vector superradiant instabilities}",
    eprint = "2407.00767",
    archivePrefix = "arXiv",
    primaryClass = "gr-qc",
    doi = "10.1103/PhysRevD.110.083029",
    journal = "Phys. Rev. D",
    volume = "110",
    number = "8",
    pages = "083029",
    year = "2024"
}

@article{Mirasola:2025car,
    author = "Mirasola, Lorenzo and others",
    title = "{Search for continuous gravitational wave signals from luminous dark photon superradiance clouds with LVK O3 observations}",
    eprint = "2501.02052",
    archivePrefix = "arXiv",
    primaryClass = "gr-qc",
    doi = "10.1103/PhysRevD.111.084032",
    journal = "Phys. Rev. D",
    volume = "111",
    number = "8",
    pages = "084032",
    year = "2025"
}

@article{Tsukada:2018mbp,
    author = "Tsukada, Leo and Callister, Thomas and Matas, Andrew and Meyers, Patrick",
    title = "{First search for a stochastic gravitational-wave background from ultralight bosons}",
    eprint = "1812.09622",
    archivePrefix = "arXiv",
    primaryClass = "astro-ph.HE",
    doi = "10.1103/PhysRevD.99.103015",
    journal = "Phys. Rev. D",
    volume = "99",
    number = "10",
    pages = "103015",
    year = "2019"
}

@article{Tsukada:2020lgt,
    author = "Tsukada, Leo and Brito, Richard and East, William E. and Siemonsen, Nils",
    title = "{Modeling and searching for a stochastic gravitational-wave background from ultralight vector bosons}",
    eprint = "2011.06995",
    archivePrefix = "arXiv",
    primaryClass = "astro-ph.HE",
    doi = "10.1103/PhysRevD.103.083005",
    journal = "Phys. Rev. D",
    volume = "103",
    number = "8",
    pages = "083005",
    year = "2021"
}

@article{Yuan:2021ebu,
    author = "Yuan, Chen and Brito, Richard and Cardoso, Vitor",
    title = "{Probing ultralight dark matter with future ground-based gravitational-wave detectors}",
    eprint = "2106.00021",
    archivePrefix = "arXiv",
    primaryClass = "gr-qc",
    doi = "10.1103/PhysRevD.104.044011",
    journal = "Phys. Rev. D",
    volume = "104",
    number = "4",
    pages = "044011",
    year = "2021"
}

@article{Yuan:2022bem,
    author = "Yuan, Chen and Jiang, Yang and Huang, Qing-Guo",
    title = "{Constraints on an ultralight scalar boson from Advanced LIGO and Advanced Virgo{\textquoteright}s first three observing runs using the stochastic gravitational-wave background}",
    eprint = "2204.03482",
    archivePrefix = "arXiv",
    primaryClass = "astro-ph.CO",
    doi = "10.1103/PhysRevD.106.023020",
    journal = "Phys. Rev. D",
    volume = "106",
    number = "2",
    pages = "023020",
    year = "2022"
}

@article{Guo:2023gfc,
    author = "Guo, Rong-Zhen and Jiang, Yang and Huang, Qing-Guo",
    title = "{Probing ultralight tensor dark matter with the stochastic gravitational-wave background from advanced LIGO and Virgo's first three observing runs}",
    eprint = "2312.16435",
    archivePrefix = "arXiv",
    primaryClass = "astro-ph.CO",
    doi = "10.1088/1475-7516/2024/04/053",
    journal = "JCAP",
    volume = "04",
    pages = "053",
    year = "2024"
}

@article{Guo:2022mpr,
    author = "Guo, Yin-da and Bao, Shou-shan and Zhang, Hong",
    title = "{Subdominant modes of the scalar superradiant instability and gravitational wave beats}",
    eprint = "2212.07186",
    archivePrefix = "arXiv",
    primaryClass = "gr-qc",
    reportNumber = "Phys. Rev. D 107, 075009 (2023)",
    doi = "10.1103/PhysRevD.107.075009",
    journal = "Phys. Rev. D",
    volume = "107",
    number = "7",
    pages = "075009",
    year = "2023"
}

@article{LIGOScientific:2014pky,
    author = "Aasi, J. and others",
    collaboration = "LIGO Scientific",
    title = "{Advanced LIGO}",
    eprint = "1411.4547",
    archivePrefix = "arXiv",
    primaryClass = "gr-qc",
    doi = "10.1088/0264-9381/32/7/074001",
    journal = "Class. Quant. Grav.",
    volume = "32",
    pages = "074001",
    year = "2015"
}

@article{Evans:2021gyd,
    author = "Evans, Matthew and others",
    title = "{A Horizon Study for Cosmic Explorer: Science, Observatories, and Community}",
    eprint = "2109.09882",
    archivePrefix = "arXiv",
    primaryClass = "astro-ph.IM",
    reportNumber = "CE-P2100003-v7, Cosmic Explorer technical report CE-P2100003-v6",
    month = "9",
    year = "2021"
}

@article{Evans:2023euw,
    author = "Evans, Matthew and others",
    title = "{Cosmic Explorer: A Submission to the NSF MPSAC ngGW Subcommittee}",
    eprint = "2306.13745",
    archivePrefix = "arXiv",
    primaryClass = "astro-ph.IM",
    month = "6",
    year = "2023"
}

@article{Punturo:2010zz,
    author = "Punturo, M. and others",
    editor = "Ricci, Fulvio",
    title = "{The Einstein Telescope: A third-generation gravitational wave observatory}",
    doi = "10.1088/0264-9381/27/19/194002",
    journal = "Class. Quant. Grav.",
    volume = "27",
    pages = "194002",
    year = "2010"
}

@article{ET:2019dnz,
    author = "Maggiore, Michele and others",
    collaboration = "ET",
    title = "{Science Case for the Einstein Telescope}",
    eprint = "1912.02622",
    archivePrefix = "arXiv",
    primaryClass = "astro-ph.CO",
    doi = "10.1088/1475-7516/2020/03/050",
    journal = "JCAP",
    volume = "03",
    pages = "050",
    year = "2020"
}

@article{LIGOScientific:2026pwx,
    author = "Abac, A. G. and others",
    collaboration = "LIGO Scientific, VIRGO, KAGRA",
    title = "{Constraints on ultralight bosons from merging binary and remnant black holes observed during the second and third parts of the fourth LIGO-Virgo-KAGRA observing run}",
    eprint = "2608.11620",
    archivePrefix = "arXiv",
    primaryClass = "gr-qc",
    reportNumber = "LIGO-P2600218",
    month = "8",
    year = "2026"
}

@article{Cardoso:2018tly,
    author = "Cardoso, Vitor and Dias, {\'O}scar J. C. and Hartnett, Gavin S. and Middleton, Matthew and Pani, Paolo and Santos, Jorge E.",
    title = "{Constraining the mass of dark photons and axion-like particles through black-hole superradiance}",
    eprint = "1801.01420",
    archivePrefix = "arXiv",
    primaryClass = "gr-qc",
    doi = "10.1088/1475-7516/2018/03/043",
    journal = "JCAP",
    volume = "03",
    pages = "043",
    year = "2018"
}

@article{Fernandez:2019qbj,
    author = "Fernandez, Nicolas and Ghalsasi, Akshay and Profumo, Stefano",
    title = "{Superradiance and the Spins of Black Holes from LIGO and X-ray binaries}",
    eprint = "1911.07862",
    archivePrefix = "arXiv",
    primaryClass = "hep-ph",
    month = "11",
    year = "2019"
}

@article{Ng:2019jsx,
    author = "Ng, Ken K. Y. and Hannuksela, Otto A. and Vitale, Salvatore and Li, Tjonnie G. F.",
    title = "{Searching for ultralight bosons within spin measurements of a population of binary black hole mergers}",
    eprint = "1908.02312",
    archivePrefix = "arXiv",
    primaryClass = "gr-qc",
    doi = "10.1103/PhysRevD.103.063010",
    journal = "Phys. Rev. D",
    volume = "103",
    number = "6",
    pages = "063010",
    year = "2021"
}

@article{Ng:2020ruv,
    author = "Ng, Ken K. Y. and Vitale, Salvatore and Hannuksela, Otto A. and Li, Tjonnie G. F.",
    title = "{Constraints on Ultralight Scalar Bosons within Black Hole Spin Measurements from the LIGO-Virgo GWTC-2}",
    eprint = "2011.06010",
    archivePrefix = "arXiv",
    primaryClass = "gr-qc",
    doi = "10.1103/PhysRevLett.126.151102",
    journal = "Phys. Rev. Lett.",
    volume = "126",
    number = "15",
    pages = "151102",
    year = "2021"
}

@article{Cheng:2022jsw,
    author = "Cheng, Lei-dong and Zhang, Hong and Bao, Shou-shan",
    title = "{Constraints on an axionlike particle from black hole spin superradiance}",
    eprint = "2201.11338",
    archivePrefix = "arXiv",
    primaryClass = "gr-qc",
    doi = "10.1103/PhysRevD.107.063021",
    journal = "Phys. Rev. D",
    volume = "107",
    number = "6",
    pages = "063021",
    year = "2023"
}

@article{Hoof:2024quk,
    author = "Hoof, Sebastian and Marsh, David J. E. and Sisk-Reyn{\'e}s, J{\'u}lia and Matthews, James H. and Reynolds, Christopher",
    title = "{Getting more out of black hole superradiance: a statistically rigorous approach to ultralight boson constraints from black hole spin measurements}",
    eprint = "2406.10337",
    archivePrefix = "arXiv",
    primaryClass = "hep-ph",
    doi = "10.1093/mnras/staf1564",
    journal = "Mon. Not. Roy. Astron. Soc.",
    volume = "546",
    number = "2",
    pages = "staf1564",
    year = "2026"
}

@article{Aswathi:2025nxa,
    author = "Aswathi, P. S. and East, William E. and Siemonsen, Nils and Sun, Ling and Jones, Dana",
    title = "{Ultralight boson constraints from gravitational wave observations of spinning binary black holes}",
    eprint = "2507.20979",
    archivePrefix = "arXiv",
    primaryClass = "gr-qc",
    doi = "10.1103/n5hr-zljn",
    journal = "Phys. Rev. D",
    volume = "112",
    number = "12",
    pages = "123048",
    year = "2025"
}

@article{Ning:2026ebu,
    author = "Ning, Orion and Safdi, Benjamin R. and Welch, Catherine",
    title = "{No Evidence for Superradiant Axions in LIGO-Virgo-KAGRA GWTC-5 Binary Black Hole Spins}",
    eprint = "2607.01317",
    archivePrefix = "arXiv",
    primaryClass = "hep-ph",
    month = "7",
    year = "2026"
}

@article{Kou:2026naz,
    author = "Kou, Xiao-Xiao and Mandic, Vuk and Ding, Ran and Tian, Chi",
    title = "{Ultralight Bosons Explain the Mass-Spin Correlations in the Merging Binary Black Hole Population}",
    eprint = "2609.02678",
    archivePrefix = "arXiv",
    primaryClass = "gr-qc",
    month = "9",
    year = "2026"
}

@article{Witte:2024drg,
    author = "Witte, Samuel J. and Mummery, Andrew",
    title = "{Stepping up superradiance constraints on axions}",
    eprint = "2412.03655",
    archivePrefix = "arXiv",
    primaryClass = "hep-ph",
    doi = "10.1103/PhysRevD.111.083044",
    journal = "Phys. Rev. D",
    volume = "111",
    number = "8",
    pages = "083044",
    year = "2025"
}

@article{Omiya:2022mwv,
    author = "Omiya, Hidetoshi and Takahashi, Takuya and Tanaka, Takahiro",
    title = "{Adiabatic evolution of the self-interacting axion field around rotating black holes}",
    eprint = "2201.04382",
    archivePrefix = "arXiv",
    primaryClass = "gr-qc",
    reportNumber = "KUNS 2916",
    doi = "10.1093/ptep/ptac058",
    journal = "PTEP",
    volume = "2022",
    number = "4",
    pages = "043E03",
    year = "2022"
}

@article{Omiya:2022gwu,
    author = "Omiya, Hidetoshi and Takahashi, Takuya and Tanaka, Takahiro and Yoshino, Hirotaka",
    title = "{Impact of multiple modes on the evolution of self-interacting axion condensate around rotating black holes}",
    eprint = "2211.01949",
    archivePrefix = "arXiv",
    primaryClass = "gr-qc",
    reportNumber = "KUNS 2945",
    doi = "10.1088/1475-7516/2023/06/016",
    journal = "JCAP",
    volume = "06",
    pages = "016",
    year = "2023"
}

@article{Boyer:1966qh,
    author = "Boyer, Robert H. and Lindquist, Richard W.",
    title = "{Maximal analytic extension of the Kerr metric}",
    doi = "10.1063/1.1705193",
    journal = "J. Math. Phys.",
    volume = "8",
    pages = "265",
    year = "1967"
}

@article{Baumann:2019eav,
    author = "Baumann, Daniel and Chia, Horng Sheng and Stout, John and ter Haar, Lotte",
    title = "{The Spectra of Gravitational Atoms}",
    eprint = "1908.10370",
    archivePrefix = "arXiv",
    primaryClass = "gr-qc",
    doi = "10.1088/1475-7516/2019/12/006",
    journal = "JCAP",
    volume = "12",
    pages = "006",
    year = "2019"
}

@article{Bao:2022hew,
    author = "Bao, Shoushan and Xu, Qixuan and Zhang, Hong",
    title = "{Improved analytic solution of black hole superradiance}",
    eprint = "2201.10941",
    archivePrefix = "arXiv",
    primaryClass = "gr-qc",
    doi = "10.1103/PhysRevD.106.064016",
    journal = "Phys. Rev. D",
    volume = "106",
    number = "6",
    pages = "064016",
    year = "2022"
}

@article{Bao:2023xna,
    author = "Bao, Shou-Shan and Xu, Qi-Xuan and Zhang, Hong",
    title = "{Next-to-leading-order solution to Kerr-Newman black hole superradiance}",
    eprint = "2301.05317",
    archivePrefix = "arXiv",
    primaryClass = "gr-qc",
    doi = "10.1103/PhysRevD.107.064037",
    journal = "Phys. Rev. D",
    volume = "107",
    number = "6",
    pages = "064037",
    year = "2023"
}

@article{Thrane:2013oya,
    author = "Thrane, Eric and Romano, Joseph D.",
    title = "{Sensitivity curves for searches for gravitational-wave backgrounds}",
    eprint = "1310.5300",
    archivePrefix = "arXiv",
    primaryClass = "astro-ph.IM",
    doi = "10.1103/PhysRevD.88.124032",
    journal = "Phys. Rev. D",
    volume = "88",
    number = "12",
    pages = "124032",
    year = "2013"
}

@article{Phinney:2001di,
    author = "Phinney, E. S.",
    title = "{A Practical theorem on gravitational wave backgrounds}",
    eprint = "astro-ph/0108028",
    archivePrefix = "arXiv",
    month = "7",
    year = "2001"
}

@article{Planck:2018vyg,
    author = "Aghanim, N. and others",
    collaboration = "Planck",
    title = "{Planck 2018 results. VI. Cosmological parameters}",
    eprint = "1807.06209",
    archivePrefix = "arXiv",
    primaryClass = "astro-ph.CO",
    doi = "10.1051/0004-6361/201833910",
    journal = "Astron. Astrophys.",
    volume = "641",
    pages = "A6",
    year = "2020",
    note = "[Erratum: Astron.Astrophys. 652, C4 (2021)]"
}

@article{Thorne:1974ve,
    author = "Thorne, Kip S.",
    title = "{Disk accretion onto a black hole. 2. Evolution of the hole.}",
    doi = "10.1086/152991",
    journal = "Astrophys. J.",
    volume = "191",
    pages = "507--520",
    year = "1974"
}

@article{LIGOScientific:2025pvj,
    author = "Abac, A. G. and others",
    collaboration = "LIGO Scientific, VIRGO, KAGRA",
    title = "{GWTC-4.0: Population Properties of Merging Compact Binaries}",
    eprint = "2508.18083",
    archivePrefix = "arXiv",
    primaryClass = "astro-ph.HE",
    reportNumber = "LIGO-P2400004",
    doi = "10.3847/2041-8213/ae771e",
    journal = "Astrophys. J. Lett.",
    volume = "1005",
    number = "2",
    pages = "L51",
    year = "2026"
}

@article{Berti:2007fi,
    author = "Berti, Emanuele and Cardoso, Vitor and Gonzalez, Jose A. and Sperhake, Ulrich and Hannam, Mark and Husa, Sascha and Bruegmann, Bernd",
    title = "{Inspiral, merger and ringdown of unequal mass black hole binaries: A Multipolar analysis}",
    eprint = "gr-qc/0703053",
    archivePrefix = "arXiv",
    doi = "10.1103/PhysRevD.76.064034",
    journal = "Phys. Rev. D",
    volume = "76",
    pages = "064034",
    year = "2007"
}

@article{Ellis:2026gwt,
    author = {Ellis, Sebastian A. R. and Ning, Orion and Rodd, Nicholas L. and Sch{\"u}tte-Engel, Jan},
    title = "{Crowdsourcing Gravitational Waves from Superradiant Axions}",
    eprint = "2603.15734",
    archivePrefix = "arXiv",
    primaryClass = "hep-ph",
    reportNumber = "RIKEN-iTHEMS-Report-26",
    month = "3",
    year = "2026"
}

@article{Scheel:2008rj,
    author = "Scheel, Mark A. and Boyle, Michael and Chu, Tony and Kidder, Lawrence E. and Matthews, Keith D. and Pfeiffer, Harald P.",
    title = "{High-accuracy waveforms for binary black hole inspiral, merger, and ringdown}",
    eprint = "0810.1767",
    archivePrefix = "arXiv",
    primaryClass = "gr-qc",
    doi = "10.1103/PhysRevD.79.024003",
    journal = "Phys. Rev. D",
    volume = "79",
    pages = "024003",
    year = "2009"
}

@article{Barausse:2009uz,
    author = "Barausse, Enrico and Rezzolla, Luciano",
    title = "{Predicting the direction of the final spin from the coalescence of two black holes}",
    eprint = "0904.2577",
    archivePrefix = "arXiv",
    primaryClass = "gr-qc",
    doi = "10.1088/0004-637X/704/1/L40",
    journal = "Astrophys. J. Lett.",
    volume = "704",
    pages = "L40--L44",
    year = "2009"
}

@article{Allen:1997ad,
    author = "Allen, Bruce and Romano, Joseph D.",
    title = "{Detecting a stochastic background of gravitational radiation: Signal processing strategies and sensitivities}",
    eprint = "gr-qc/9710117",
    archivePrefix = "arXiv",
    reportNumber = "WISC-MILW-97-TH-14",
    doi = "10.1103/PhysRevD.59.102001",
    journal = "Phys. Rev. D",
    volume = "59",
    pages = "102001",
    year = "1999"
}

@article{Lyu:2026mbv,
    author = "Lyu, Zhen-Hong and Cai, Rong-Gen and Liu, Jing",
    title = "{Self-stimulated avalanche in superradiant axion clouds}",
    eprint = "2606.10776",
    archivePrefix = "arXiv",
    primaryClass = "gr-qc",
    month = "6",
    year = "2026"
}
\end{document}